\documentclass[final]{siamltex}

\usepackage{amsmath,amssymb,latexsym}

\usepackage[dvips]{graphicx}
\usepackage{subfigure}
\usepackage{psfrag}
\usepackage[usenames]{color}

\newcommand{\rbd}[1]{\raisebox{-0.5ex}[0.5ex]{#1}}
\newcommand{\rbu}[1]{\raisebox{0.5ex}[-0.5ex]{#1}}
\newcommand{\rb}[1]{\raisebox{2ex}[-2ex]{#1}}
\newfont{\twelvebf}{cmbx12}
\newfont{\twelverm}{cmr12}
\newfont{\twelveit}{cmti12}

\newcommand{\parmetis}{{\sc{ParMeTiS}}}

\newcommand{\bv}{\mathbf{v}}

\newcommand{\ba}{\mathbf{a}}

\newcommand{\sg}{\sqrt{G}}
\newcommand{\citep}{\cite}

\title{A Comparison of Two Shallow Water Models with Non-Conforming Adaptive Grids: classical tests}

\author{
Amik St-Cyr\thanks{Corresponding author address: National Center for Atmospheric 
Research (NCAR), 1850 Table Mesa Drive Boulder, CO 80305, USA, E-mail: amik@ucar.edu}
\and 
Christiane Jablonowski\thanks{Current affiliation: University of Michigan, Department of Atmospheric, Oceanic and Space Sciences, Ann Arbor, MI, USA}
\and 
John M. Dennis\thanks{National Center for Atmospheric Research (NCAR), 1850 Table Mesa Drive Boulder, CO 80305, USA} 
\and 
Henry Tufo\thanks{Second affiliation: University of Colorado Boulder, 430 UCB Boulder,  CO 80309-0430, USA} 
 \and  Stephen J. Thomas\thanks{National Center for Atmospheric Research (NCAR), 1850 Table Mesa Drive Boulder, CO 80305, USA}}
\begin{document}
\maketitle

\begin{abstract} 
In an effort to study the applicability of adaptive mesh refinement (AMR) techniques to atmospheric 
models an interpolation-based spectral element 
shallow water model on a cubed-sphere grid is compared to a block-structured 
finite volume method in latitude-longitude geometry. Both models utilize a non-conforming
adaptation approach which doubles the resolution at fine-coarse mesh interfaces. The underlying AMR libraries
are quad-tree based and ensure that neighboring regions can only differ by one refinement level.
The models are compared via selected test cases from a standard test suite for the shallow water equations.
They include the advection of a cosine bell, a steady-state 
geostrophic flow, a flow over an idealized  mountain and a Rossby-Haurwitz wave. Both static and dynamics adaptations
are evaluated which reveal the strengths and weaknesses of the AMR techniques.
Overall, the AMR simulations show that both models successfully place static and dynamic adaptations in
local regions without requiring a fine grid in the global domain. 
The adaptive grids reliably track features of interests without visible distortions or noise
at mesh interfaces. Simple threshold adaptation criteria for the geopotential height and the relative vorticity are
assessed.
\end{abstract}

\section{Introduction}
Simulating the climate is a grand challenge problem requiring multiple,
century-long integrations of the equations modeling  the Earth's 
atmosphere. Consequently, grid resolutions in atmospheric climate 
models are much coarser than in numerical weather models,
where accurate predictions are limited to about ten days. However,
it has been recognized that localized flow structures, like hurricanes,  may 
play an important role in obtaining the correct climate signal.
In particular, there exists a requirement for localized mesh refinement
in regional climate modeling studies. It is also conjectured, at this time, 
that the lack of resolution in the tropics is the cause of the inability of 
most climate models to capture the statistical contribution of extreme 
weather events.

While mesh adaptation is a mature field in computational fluid dynamics, 
currently, the only fully operational adaptive weather and dispersion model is 
OMEGA \citep{bacon:2000,Gopalakrishnan:2002in}. The latter is based on the 
finite-volume approach and uses conforming Delaunay meshes that are locally 
modified and smoothed. Researchers involved in hurricane or typhoon 
predictions where amongst the very first to experiment with variable resolutions 
in atmospheric models. The preferred method consisted 
of nesting finer meshes statically into coarser ones \citep{Kurihara:1979zv,Kurihara:1980jj}. 
Nesting is a wide spread technique in current models, especially mesoscale models, 
to achieve high resolutions \citep{Frohn:2002zp} in local regions.  Truly adaptive models 
were first developed by \cite{Skamarock:1989ya}, \cite{Skamarock:1993mk} and 
\cite{Dietachmayer:1992tg}. \cite{Skamarock:1993mk}  based their adaptive mesh refinement (AMR)
strategy on a truncation error estimate \citep{Skamarock:1989gy}.  Continuing in the 
realm of second order conforming methods \cite{Iselin:2002ff} developed a dynamically 
adaptive Multidimensional Positive Definite Advection Transport Algorithm (MPDATA) 
\citep{Iselin:2005ma}. MPDATA  was thoroughly reviewed in 
\cite{Smolarkiewicz:2006ko} because of its various numerical qualities. The discussion
addressed generalized space-time coordinates that
enable continuous mesh deformations  \citep{Prusa:2003vs}  as well as the generalization of MPDATA to 
unstructured grids \citep{Smolarkiewicz:2005ih}. Recently, adaptive schemes on the sphere 
were also discussed by \cite{Jablonowski:2004fa}, \cite{laeuter:2004}, \cite{behrens:2005} and
\cite{laeuter:2006} where only \cite{Jablonowski:2004} utilizes a 
non-conforming approach. A review of adaptive methods in atmospheric 
modeling is given in \cite{behrens:2006}.

It is recognized that high-order spatial resolution is necessary in climate 
modeling. To paraphrase \cite{Boyd:2004lc}, spectral methods like continuous 
Galerkin (CG) or discontinuous Galerkin (DG) are blessedly free of the spurious wave 
dispersion induced by low-order methods\footnote{The quote actually starts with:  
"It is all about the waves stupid: ..."}.  The corresponding reduction in grid points directly 
lowers the amount of costly column physics evaluations. Multiple efforts
to include conforming mesh refinement techniques in models with high-order numerical methods  were initiated 
\cite{Giraldo:2000nu},  \cite{Fournier:2004wk}, \cite{Giraldo:2005yj}, \cite{Giraldo:2006gw}
and \cite{Rosenberg:2006zs}. However, supporting conforming adaptive meshes has negative 
effects on the Courant-Friedrich-Levy (CFL) stability condition. One solution is to consider 
non-conforming elements or blocks.

In this paper two adaptive mesh techniques for 2D shallow water flows on the sphere 
are compared. In particular, the study focuses on the characteristics of the AMR 
approach in the  cubed-sphere spectral element model (SEM) by \cite{St-Cyr:2006xf} and 
compares it to the adaptive finite volume (FV) method by \cite{Jablonowski:2006}. 
Both SEM and FV are AMR models of the non-conforming type. 
A thorough study of the dynamically adaptive FV scheme based 
on quadrilateral control volumes can also be found in \cite{Jablonowski:2004fa} and \cite{Jablonowski:2006b}. 
In particular, 2D shallow water and 3D primitive-equation numerical experiments were 
conducted using a latitude-longitude grid on the sphere. This adaptive model is built 
upon the finite volume technique by \cite{Lin:1996} and \cite{Lin:1997}. The spectral 
element model is based on the ideas of \cite{Patera:1984uq}. The non-conforming 
treatment follows the interpolation procedure by \cite{Fischer:2002sx}. 
The spectral element method was originally developed for incompressible fluid flows. 
Meanwhile, it has been adapted by 
many authors \citep{Haidvogel:1997kx,Taylor:1997vn,Giraldo:2001ys,Thomas:2002ti} 
for global atmospheric and oceanic simulations.

The paper is organized as follows. In Section \ref{sec:sw} the shallow water 
equations are introduced which are the underlying equation for our model comparison. 
Both models SEM and FV are briefly reviewed in Section~\ref{sec:sw}. This includes a 
discussion of the adaptive mesh approach for the spectral elements in SEM and the 
latitude-longitude blocks in FV. In Section~\ref{sec:results} the characteristics of the 
AMR techniques are tested using selected test cases from the \cite{Williamson:1992ts} 
shallow water test suite. They include the advection of a cosine bell, a steady-state 
geostrophic flow, a flow over a mountain and a Rossby-Haurwitz wave. The findings are
summarized in Section~\ref{sec:conclusion}.

%------------------------------------------------------------------------------------------
\section{Shallow water equations}
\label{sec:sw}
The shallow water equations have been used as a test bed for
promising numerical methods by the atmospheric
modeling community for many years. They contain the essential wave propagation
mechanisms found in atmospheric General Circulation Models (GCMs). The 
linearized primitive equations yield a series of layered 
shallow water problems where the mean depth of each layer is related 
to the maximum wave speed supported by the medium.
These are the fast-moving gravity waves and nonlinear Rossby waves.  
The latter are important for correctly capturing nonlinear atmospheric dynamics. 
The governing equations of motion for the inviscid flow of a free surface are given by
\begin{align}
\label{eq:3.1} 
\frac{\partial\,\bv}{\partial t} + 
(f + \zeta)\:{\bf k}\times\bv + \frac{1}{2} \mathbf{\nabla} \left( \bv\cdot\bv \right) 
+ \mathbf{\nabla} (\Phi + \Phi_s) & = 0 \\
\frac{\partial\,\Phi}{\partial t} + \mathbf{\nabla\cdot}(\Phi \bv) & =  0 \label{eq:3.2}
\end{align}
where  $t$ is the time, $\Phi=g h$ symbolizes the geopotential, $h$ is the height above sea 
level and $g$ denotes the gravitational acceleration. Furthermore, $\Phi_s = g h_s$ is the 
surface geopotential, $h_s$ symbolizes the height of the orography,  $\bv$ stands for the 
horizontal velocity vector,  ${\bf k}$ denotes the radial outward unit vector, and $f$ 
and $\zeta$ are the Coriolis parameter and relative vorticity, respectively. $\mathbf{\nabla}$ 
indicates the horizontal gradient operator, whereas $\mathbf{\nabla \cdot}$ stands for 
the horizontal divergence operator. The momentum equation (\ref{eq:3.1}) is written in  
its vector-invariant form.
\section{Description of the adaptive shallow water models}
\label{sec:AMR_sw}
The spectral element method (SEM) is a combination of ideas coming from the 
finite-element method and of the spectral method.  The order of accuracy is determined 
by the degree of the local basis functions within a finite element. The basis consists of 
Lagrange polynomials passing through Gauss-Legendre-Lobatto (GLL) or Gauss-Legendre (GL)
quadrature points facilitating greatly the evaluation of the integrals appearing in a weak 
formulation. For the tests here, fifth degree basis functions for scalars are employed which 
utilize $6 \times 6$ GL quadrature points per spectral element. 
The SEM grid is based on a projection of a cube inscribed into a 
sphere (\cite{Sadourny:1972xo}), a so-called cubed-sphere grid
which consists of 6 faces. These are further subdivided into
spectral elements quadrangles which are evenly distributed over the surface of 
the sphere in the non-adapted case. Note that the numerical approach in SEM is 
non-monotonic and does not conserve mass. Nevertheless, the variation in the 
total mass is small over typical forecast periods of two weeks.

This is in contrast to the monotonic and mass-conservative FV model which was originally developed 
by \cite{Lin:1997}. It utilizes the third order Piecewise Parabolic Method (PPM) 
\citep{Colella:1984hi} which was first designed for compressible fluids  with strong 
shocks. The PPM algorithm applies monotonicity constraints that act as a nonlinear 
scale-selective dissipation mechanism. This dissipation primarily targets the flow features at the
smallest scales.

Both shallow water models are characterized in more detail below. In particular, the 
individual adaptive mesh approaches are described which implement adaptations in the
horizontal directions. The time step, on the other hand, is not adapted, 
except for selected advection 
tests. Therefore, the chosen time step must be numerically stable on the finest grid in an 
adapted model run.

\subsection{Spectral element (SEM) shallow water model}
\label{subsec:sem}
\subsubsection{Curvilinear coordinates: cubed sphere}
The flux form shallow-water equations in curvilinear coordinates
are described in \cite{Sadourny:1972xo}.  Let $\ba_1$ and $\ba_2$ be
the covariant base vectors of the transformation between
inscribed cube and spherical surface. The metric tensor of the 
transformation is defined as
$G_{ij} \equiv \ba_i \cdot \ba_j$.
Covariant and contravariant vectors are related through the metric 
tensor by $u_i = G_{ij} u^j$ and
$u^i = G^{ij} u_j$,  where $G^{ij} = (G_{ij})^{-1}$  and $G = \det (G_{ij})$.
The six local coordinate systems $(x^1, x^2)$ are based on an equiangular 
central projection, $-\pi/4 \leq x^1, x^2 \leq \pi/4$. The metric tensor for 
all six faces of the cube is
\begin{equation}
  G_{ij} = \frac{1}{r^4 \cos^2 x^1 \, \cos^2 x^2} \,
    \left[   \begin{array}{cc}
                1 + \tan^2 x^1    & -\tan x^1 \, \tan x^2 \\
               - \tan x^1 \, \tan x^2  & 1 + \tan^2 x^2
             \end{array}
   \right]           \label{mgij}
\end{equation}
where $r = (1 + \tan^2 x^1 + \tan^2 x^2)^{1/2}$
and $\sg = 1/r^3 \, \cos^2 x^1 \, \cos^2 x^2 $. 

The shallow water equations are written in curvilinear coordinates using the 
following definitions for divergence and vorticity 
\begin{equation*}
\nabla\cdot{\bf v} \equiv \frac{1}{\sqrt{G}}
\left[
\frac{\partial}{\partial x^i}\;(\;\sqrt{G}\;u^i\;)
\right],
\quad
\zeta \equiv \frac{\epsilon_{ij}}{\sqrt{G}}\frac{\partial u_j}{\partial x^i} =
\frac{1}{\sqrt{G}}\left[ 
\frac{\partial u_2}{\partial x^1} -
\frac{\partial u_1}{\partial x^2} \right] 
\label{eq:3.4}
\end{equation*}
and replacing them in (\ref{eq:3.2}). In their contravariant form the equations are
\begin{align}
\frac{\partial u^i}{\partial t} + G^{ij} \left[ \frac{\partial}{\partial x^j} ( u^k u_k)
  +\frac{\partial\Phi}{\partial x^j}+\sg \, \epsilon_{jk}u^k (f + \zeta)\right]&=0,   \label{mom1} \\
\frac{\partial \Phi}{\partial t} + u^j \frac{\partial\Phi}{\partial x^j}+ \frac{\Phi}{\sqrt{G}}\frac{\partial}{\partial x^j}(\sqrt{G} u^j) & =   0   \label{conti} 
\end{align}
where the Einstein summation convention is used for repeated indices.
Additional information on the equation set can also be found in \cite{Nair:2005aj}.

\subsubsection{Spatial and temporal discretization}
\begin{figure}
  \centering
  \includegraphics[scale=0.35]{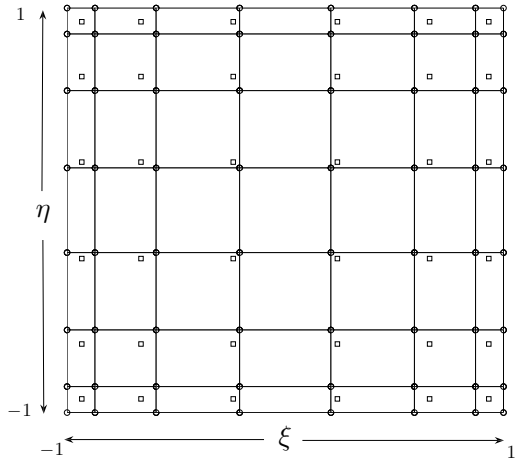}
  \caption{Positions of the GLL (circles) and GL (open squares) quadrature points on the spectral
  reference element with mapped coordinates $\xi$ and $\eta$. The polynomial degrees are 7 
  (GLL velocity points) and 5 (GL scalar points).}
  \label{fig:sem_gridpoints}
\end{figure}

The equations of motion are discretized in space using the 
${\mathbb P}_N - {\mathbb P}_{N-2}$ spectral element method as in 
\cite{Thomas:2002ti}. The cubed-sphere is partitioned into 
$K$ elements $\Omega^k$ in which the dependent and 
independent variables are approximated by tensor-product 
polynomial expansions. The velocity on element $\Omega^k$ is expanded in terms of 
the $N$-th degree Lagrangian interpolants $h_i$ 
\begin{equation}
{\bf v}_h^k({\bf x}) =
\sum_{i=0}^{N}\sum_{j=0}^{N}\: {\bf v}^k_{ij}\:h_i(\xi^k({\bf x}))\:h_j(\eta^k({\bf x}))
\label{eq:3.5}
\end{equation}
and the geopotential is expanded on the same element using the $(N-2)$--th degree
interpolants $\tilde{h}_i$
\begin{equation}
{\Phi}_h^k({\bf x}) =
\sum_{i=1}^{N-1}\sum_{j=1}^{N-1}\:
{\Phi}^k_{ij}\:\tilde{h}_i(\xi^k({\bf x}))\:\tilde{h}_j(\eta^k({\bf x}))
\end{equation}
where ${\bf x}\rightarrow (\xi^k({\bf x}),\eta^k({\bf x}))$ is an affine transformation 
from the element $\Omega^k$ on the cubed sphere to the reference element $[-1,1]\times[-1,1]$ 
pictured in Fig.~\ref{fig:sem_gridpoints}. 
A weak Galerkin formulation results from the integration of the
equations with respect to test functions and the direct evaluation
of inner products using Gauss-Legendre and Gauss-Lobatto-Legendre 
quadrature.  The positions of the corresponding GL and GLL
quadrature points within each spectral element with mapped
coordinates $\xi$ and $\eta$ are depicted in Fig.~\ref{fig:sem_gridpoints}.
The GLL points for the co-located velocity components are marked by the
circles, the open squares point to the GL points for scalars. The polynomial
degrees are 7 and 5, respectively. An overview of the 
chosen parameters and base resolutions for the present study is also given in 
Table~\ref{tab:sem_resolution}.

\begin{table}[htb]
\caption {Overview of the resolutions in SEM.}
\begin{center}
\begin{tabular}{ccccc} \hline \hline
\rbd{\# Elements}      & \rbd{Total \# of}  & \rbd{\# GL points} & \rbd{\# GLL points} &  \rbd{Approximate}  \\
\rbu{per cubed face} &  \rbu{elements} & \rbu{per element}   & \rbu{per element}  & \rbu{resolution}       \\ \hline
$3 \times 3$ & 54   & $6 \times 6$ & $8 \times 8$ &  $5^{\circ} \times 5^{\circ}$ \\
$4 \times 4$ & 96   & $6 \times 6$ & $8 \times 8$ & $3.2^{\circ} \times 3.2^{\circ}$ \\
$6 \times 6$ & 216 & $6 \times 6$ & $8 \times 8$ & $2.5^{\circ} \times 2.5^{\circ}$ \\ \hline
\end{tabular}
\end{center}
\label{tab:sem_resolution}
\end{table}

$C^0$ continuity of the velocity is enforced at inter-element
boundaries sharing Gauss-Lobatto-Legendre points and direct stiffness
summation (DSS) is then applied \citep{Tufo:1998ey,Deville:2002fn}.
The advection operator in the momentum equation is expressed
in terms of the relative vorticity and kinetic energy, whereas the
continuity equation relies on the velocity form. 
\cite{Wilhelm:uo} have shown that the rotational form of 
the advection operator is stable for the ${\mathbb P}_N - {\mathbb P}_{N-2}$ 
spectral element discretization. A standard Asselin-Robert filtered 
leap-frog discretization is employed for integrating the equations of 
motion in time. More advanced options, that are tailored for AMR
based on the ideas in \cite{St-Cyr:2005ur}, will be described in \cite{St-Cyr:2006xf}. 
Since a spectral basis is employed on each element, the spectrum of the 
advective operators must be corrected using a $2/3$ rule similar to what is employed 
in global pseudo spectral models. A Boyd-Vandeven filter \citep{Vandeven:1991ef,Boyd:1996hn}
was used during the adaptive numerical 
simulations that removes the last third of the spectrum.
\subsubsection{Non-conforming spectral elements}
The positions of the collocation points at the boundaries of non-conforming spectral elements 
do not coincide and a procedure is required to connect neighboring elements.
Several techniques are available including mortars \citep{Mavriplis:1989xe,Feng:2002uq}  
and interpolations \citep{Fischer:2002sx}. Here, interpolations are used.
The true unknowns at a boundary belong 
to the master coarse element and are passed to the refined slave elements by the following
procedure. From Eq.~(\ref{eq:3.5}), at one of the element 
boundaries $\Gamma_{km} = \Omega_k \cap \Omega_m$ with $\xi^k = \xi^m =1$, the trace of the solution is 
\begin{align}
{\bf v}({\bf x}(\xi^k,\eta^k))_h^k|_{\Gamma_{km}} = {\bf v}_h^k({\bf x}(1,\eta^k))=  \sum^{N}_{j=0} {\bf v}^k_{Nj} h_j( \eta^k )
\end{align}
where $h_j$ are the Lagrange interpolants defined at the GLL points. To interpolate the master 
solution to the slave edge, a mapping is created from the master reference element relative to the 
slave's reference element.  Let $\eta^l_m$ denote this mapping, then
\begin{align}
\eta^{l}_{m}(\eta) := \frac{\eta}{2^{l}} + \frac{2m - 1}{2^{l}} - 1
\end{align}
where $m \in [1,2^{l} ]$ is the slave face number and $l$ is
the number of the refinement level. This spatial refinement strategy is
called $h$--refinement which is in contrast to increasing the
order of the polynomials ($p$--refinement).

Interpolations can be expressed in matrix form as
\begin{align}
\left[J^{l}_{m}\right]_{ij} = h_{j}(\eta^{l}_{m}(\xi_{i}))
\label{WD_INTERPOLATION}
\end{align}
where $\{\xi_{i}\}^N_{i=0}$ are the Gauss-Legendre-Lobatto  points 
used in the quadrature and in the collocation of the dependent 
variables.
If ${\bf v}^k$ are the unknowns on the edge of the master element, then 
$J^{l}_{m} {\bf v}^k$ represents the  master element contributions passed 
to the slave elements. Assembly of the global spectral element matrix is not 
viable on today's computer architectures with distributed memory. Instead, the action of the assembled matrix on a vector 
is performed with the help of DSS. 
For $A$, a matrix resulting from the spectral element discretization of a differential operator, 
defined for the true degrees of freedom and $A_L$, the block diagonal matrix of the 
individual local unassembled contributions of each element as if they were disjoint,  
the DSS for conforming elements is represented by  
\begin{align}
v^{T} A u = v^{T} Q^{T}  A_{L} Q u 
\end{align} 
where $Q$ is a Boolean rectangular matrix representing the scattering of 
the true degrees of freedom to the unassembled blocks (one block in $A_L$ per element). 
A non-conforming formulation can be obtained by replacing the scattering Boolean matrices 
$Q$ with $Q=J_{L} \tilde{Q}$
\begin{align}
v^{T} A u = v^{T} (\tilde{Q}^{T} J^{T}_{L}) A_{L} (J_{L} \tilde{Q}) u \mbox{.}
\end{align} 
The block diagonal matrix $J_L$ consist of the identity where the interfaces between elements match
and of the interpolation matrix (\ref{WD_INTERPOLATION}) where 
interfaces are non-conforming.  To facilitate time-stepping procedures, the matrix must 
be lumped by summing rows in the mass matrix \citep{Quarteroni:1997kw}.  
Let $L(A)$ represent the lumping operation. Lumping the global matrix $A$ 
is equivalent to lumping the local matrix $J^{T}_{L} A_L J_{L}$ and it can be shown that
\begin{align}
L(A) = L(\tilde{Q}^T J^{T}_{L} A_L J_{L} \tilde{Q}) = \tilde{Q}^T L(J^{T}_{L} A_L J_{L})\tilde{Q}.
\end{align}
\subsubsection{Adaptation principle}
The adaptation algorithm developed here is parallel and usable on distributed 
memory computers.  The cubed sphere is initially tiled with an uniform, low resolution, 
mesh and each element represents the root of a quad-tree. 
While a quad-tree is a natural description for 2-dimensional mesh refinement, the six 
faces of the cubed-sphere geometry necessitates a graph data structure. 
For the quad-trees,  a set of simple bit manipulation subroutines are used to
provide inheritance information.  A graph data structure which describes the 
connection between the roots and leaves of each quad-tree is also maintained.  
To simplify the interface management, neighboring elements are restricted 
to be at most one level of refinement apart.  When elements are marked for 
refinement, a parallel procedure verifies that a compatible quad-tree refinement 
exists, with respect to all the roots and the interface restriction. The communication 
package is properly modified to enable correct inter-element communications related 
to the DSS procedure. This was greatly simplified by replacing the graph-based 
communication package in the model 
by the more generic one developed by \cite{Tufo:1998ey}. If a load imbalance is 
detected then the partitioning algorithm 
is invoked and elements are migrated to rebalance the workload on each processor.  
Apart from the Message Passing Interface (MPI) library, the graph partitioning tool \parmetis 
\citep{Karypis:2002yo} and the generic DSS  libraries, no other specialized high-level 
library is used. The non-adapted spectral element model also supports an efficient
space-filling curve loadbalancing approach  \citep{dennis:03} which we plan to use in the adaptive model
for  future simulations.

\subsection{Finite volume (FV) shallow water model}
\label{subsec:fv}
\subsubsection{Model description}
The adaptive finite volume model is built upon the advection algorithm by \cite{Lin:1996} 
and its corresponding shallow water system \citep{Lin:1997}.  The FV shallow water model 
is comprised of the momentum equation and mass continuity equation as shown in 
Eqns.~(\ref{eq:3.1}) and (\ref{eq:3.2}). Here the flux-form of the mass conservation law 
and the vector-invariant form of the momentum equation are selected. In addition, a 
divergence damping term is added to the momentum equation.

The finite-volume dynamical core utilizes a flux form algorithm for the horizontal 
advection processes, which, from the physical point of view, can be considered
a discrete representation of the conservation law in finite-volume space. 
However, from the mathematical standpoint, it can be viewed as a numerical 
method for solving the governing equations in integral form. This leads to a more natural 
and often more precise representation of the advection processes, especially in 
comparison to finite difference techniques. The transport processes, e.g. for the height 
$h = \Phi/g$ of the shallow water system, are modeled by fluxes into and out of the finite 
control-volume where volume-mean quantities are predicted
\begin{eqnarray}
\bar{h}_{i,j}^{n+1} &=& \bar{h}_{i,j}^{n} -  \frac{\Delta t}{a \,\cos\phi_j \, \Delta\lambda_i} \:
\big( F_{i+\frac{1}{2},j} - F_{i-\frac{1}{2},j} \big)  \nonumber\\
& & \quad \;\,\, - \frac{\Delta t}{a \,\cos\phi_j \, \Delta\phi_j} \:
\big( \cos\phi_{j+\frac{1}{2}}\, G_{i,j+\frac{1}{2}} \,-\, \cos\phi_{j-\frac{1}{2}}\,G_{i,j-\frac{1}{2}} \big) 
\mbox{.}
\label{eqn:eqn_flux_F_G_spherical_lr}
\end{eqnarray}
Here, the overbar $\bar{( )}$ indicates the volume-mean, $F$ and $G$ denote the 
time-averaged 1D numerical fluxes in longitudinal and latitudinal direction which are 
computed with the upstream-biased and monotonic PPM scheme (see also 
\cite{carpenter:90} and \cite{nair:02}). $\Delta t$ symbolizes the time step, $a$ stands 
for the radius of the Earth, the indices $i$ and $j$ point to grid point locations in the 
longitudinal ($\lambda$) and latitudinal ($\phi$) direction and the index $n$ marks the 
discrete time level. In addition, 
$\Delta\lambda_i = (\lambda_{i+\frac{1}{2}} - \lambda_{i-\frac{1}{2}})$ and 
$\Delta\phi_i = (\phi_{j+\frac{1}{2}} - \phi_{j-\frac{1}{2}})$ represent the longitudinal 
and latitudinal grid distances measured in radians. 

The advection algorithm shown in Eq.~(\ref{eqn:eqn_flux_F_G_spherical_lr}) is the 
fundamental building block of the horizontal discretization in the FV shallow water model.
It is not only used to predict the time evolution of the mass (Eq.~(\ref{eq:3.2})), but also 
determines the absolute vorticity fluxes and kinetic energy in the momentum equation.
Further algorithmic details of the shallow water code are described in \cite{Lin:1997} 
who also discussed the staggered Arakawa C and D grid approach utilized in the model.

\subsubsection{Block-structured adaptive mesh approach}
The AMR design of the FV shallow water model is built upon a 2D block-structured data 
configuration in spherical coordinates. The concept of the block data structure and corresponding
resolutions is fully described in \cite{Jablonowski:2006}. In essence, a regular latitude-longitude grid is split 
into $B_y \times B_x$ non-overlapping spherical blocks that span the entire sphere. Each 
block is logically rectangular and comprises a constant number of $N_y \times N_x$ grid 
points in latitudinal and longitudinal direction. If, for example, $6 \times 9$ grid points per 
block are selected with $B_y \times B_x = 6 \times 8$ blocks on the sphere then the 
configuration corresponds to a $5^{\circ} \times 5^{\circ}$ uniform mesh resolution. Here, 
we use such an initial configuration for selected test cases in Section~\ref{sec:results}. All 
blocks are self-similar and split into four in the event of refinement requests, thereby doubling 
the spatial resolution. Coarsenings reverse this refinement principle. Then four ``children'' 
are coalesced into a single self-similar parent block which reduces the grid resolution in 
each direction by a factor of 2. As in SEM neighboring blocks can only differ by one 
refinement level. This leads to cascading refinement areas around the finest mesh region. 
In addition, the blocks adjacent to the poles are kept at the same refinement level due to 
the application of a fast Fourier transform (FFT) filter in longitudinal direction for nonlinear 
flows.

The FV model supports static and dynamic adaptations which are managed by an adaptive 
block-structured grid library for parallel processors \citep{oehmke:01,oehmke:04}. This 
library utilizes a quad-tree representation of the adapted grid which is similar to the 
adaptive mesh technique in SEM. The communication on parallel computer architectures 
is performed via MPI. Our load-balancing strategy aims at assigning an equal workload to 
each processor which is equivalent to an equal number of blocks in the FV shallow water model. 
No attempt is made to keep geographical regions on the same processor. This can be 
achieved by a space-filling curve load-balancing strategy as in \cite{dennis:03} which reduces 
the communication overhead. Such an approach is the subject of future research. The 
refinement criteria are user selected. In particular, simple geopotential height thresholds 
or vorticity-based criteria are assessed for our dynamic adaptation tests. For static adaptations 
though, we place the fine-grid nests in pre-determined regions like mountainous areas or 
geographical patches of interest. 

Each block is surrounded by ghost cell regions that share the information along 
adjacent block interfaces. This makes each block independent of its neighbors since the 
solution technique can now be individually applied to each block. The ghost cell information 
ensures that the requirements for the numerical stencils are satisfied. The 
algorithm then loops over all available blocks on the sphere before a communication step with 
ghost cell exchanges becomes necessary. The number of required ghost cells highly depends 
on the numerical scheme. Here, three ghost cells in all horizontal directions are needed 
which are kept at the same resolution as the inner domain of the block. The consequent interpolation 
and averaging techniques are further described in \cite{Jablonowski:2004fa} and 
\cite{Jablonowski:2006}.

The time-stepping scheme is explicit and stable for zonal and meridional Courant numbers 
$|$CFL$| < 1$. This restriction arises since the semi-Lagrangian extension of the 
\cite{Lin:1996} advection algorithm is not utilized in the AMR model experiments. This keeps 
the width of the ghost cell regions small, but on the other hand requires small time steps if 
high wind speeds are present in polar regions. Then the CFL condition is most restrictive due 
to the convergence of the meridians in the latitude-longitude grid. Therefore, we restrict 
the adaptations near the pole to very few (1-2) refinement levels.

\begin{figure}
  \centering
  \includegraphics{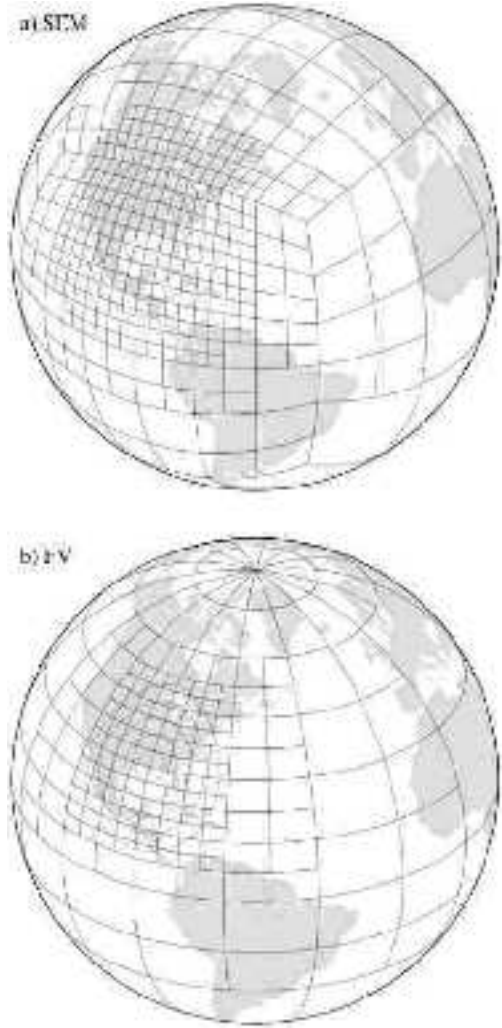}
  \caption{Adaptive grids on the sphere in (a) SEM and (b) FV. The adapted elements on the cubed-sphere 
  (SEM) and adapted blocks in latitude-longitude geometry (FV) refine an idealized mountain as indicated by the contour lines.}
  \label{fig:amr_grids}
\end{figure}

The adaptive grids of both models are compared in Fig.~\ref{fig:amr_grids}. Here, an idealized mountain as
indicated by the contour lines is refined at the maximum refinement level of 3. This corresponds to
the grid resolution $0.625^{\circ} \times 0.625^{\circ}$ in the finest adaptation region. The adaptive elements (SEM)
and blocks (FV) overlay the sphere. The differences between
the two adaptive meshes are clearly visible. The SEM model (Fig.~\ref{fig:amr_grids}a) utilizes a non-orthogonal
cubed-sphere geometry and places the refined spectral elements across an edge of two cubed-sphere faces. The FV
model (Fig.~\ref{fig:amr_grids}b) is formulated on a latitude-longitude grid. This leads to orthogonal blocks which are
closely spaced in polar regions.

\section{Numerical experiments}
\label{sec:results}
For our AMR model intercomparison, we select test cases with increasing complexity from the
standard \cite{Williamson:1992ts} shallow water test suite.
The \cite{Williamson:1992ts} test suite assesses scenarios that are
mainly characterized by large-scale smooth flow fields. Among them are
the passive advection of a cosine bell (test case 1), a steady-state geostrophic flow at 
various rotation angles (test case 2), a flow over an idealized mountain (test case 5) and a Rossby-Haurwitz 
wave with wavenumber 4 (test case 6). Our discussion of the adaptive models is focused on these tests.
Both static and dynamic refinement areas are addressed that highlight the strengths and
weaknesses of the two AMR approaches. 

Some basic statistics of the adaptive simulations
can be found in Table~\ref{tab:statistics}. The table lists not only the test case numbers with corresponding 
rotation angles $\alpha$ and base resolutions, but also gives information on the AMR approach, the number 
of refinement levels, the time steps as well as the number of adapted elements (SEM) and blocks (FV)
at certain snapshots in time. Note that the time steps are not optimized for performance. Rather they are
selected to guarantee numerical stability.
\begin{table}[htb]
\caption {Statistics for the adaptive SEM and FV simulations.}
\begin{center}
{\tiny
\begin{tabular}{cccccccccc} \hline \hline
 \rbd{Test} &                           & \rbd{Base} & \rbd{\# Ref.}   &                  & \multicolumn{2}{c}{SEM}  & & \multicolumn{2}{c}{FV}  \\ \cline{6-7} \cline{9-10}
\rbu{case} &  \rb{$\alpha$} & \rbu{resolution}  & \rbu{level} & \rb{AMR} & $\Delta t$ (s) & \# Elements &  &$\Delta t$ (s)   &   \# Blocks \\ \hline
1 & $45^{\circ}$ & $5^{\circ} \times 5^{\circ}$ & 3 & dynamic & 10 & 322 (day 9) & &adaptive & 201 (day 9) \\
1 & $90^{\circ}$ & $5^{\circ} \times 5^{\circ}$ & 3 & dynamic & 10 & 242 (day 3) & &adaptive & 480 (day 3) \\
2 & $45^{\circ}$ & $2.5^{\circ} \times 2.5^{\circ}$ & 2 & static  & 10 & 324  & &200 & 288 \\
5 & & $5^{\circ} \times 5^{\circ}$ & 3 & dynamic & 20 & 1448 (day 15) & &138 & 744 (day 15) \\
6 & & $2.5^{\circ} \times 2.5^{\circ}$ & 1 & static  & 15 & 312 & &225 & 288 \\
6 & & $3.2^{\circ} \times 3.2^{\circ}$ & 1 & static  & 15 & 176 & & -- & -- \\ \hline
\end{tabular}}
\end{center}
\label{tab:statistics}
\end{table}
The model results are typically evaluated via normalized $l_1$, $l_2$ and $l_{\infty}$ 
error norms. The definitions of the norms are given in \cite{Williamson:1992ts}. For the 
adaptive FV shallow water model the computation of the norms is also explained in 
\cite{Jablonowski:2006}.

\subsection{Passive advection of a cosine bell}
\label{subsec:test1}
We start our intercomparison of the two AMR approaches with the traditional solid body 
rotation of a cosine bell around the sphere (see test case 1 in \cite{Williamson:1992ts} for 
the initial conditions). This test evaluates the advective component of the numerical 
method in isolation. It challenges the spectral element method of SEM due to possible 
over- and undershoots of the transported feature. They are due to the so-called Gibbs 
phenomenon which is characteristic for spectral approaches \citep{Gibbs:1899fk,
Vandeven:1991ef,Boyd:1996hn}. Several values for the rotation angle $\alpha$ can be 
specified. It determines the angle between the axis of the solid body rotation and the poles 
in the spherical coordinate system. Here, we mainly report results for $\alpha=45^{\circ}$ 
which advects the cosine bell slantwise over four corner points and two edges of the 
cubed-sphere. In addition, results for $\alpha=0^{\circ}$ (transport along the equator) and 
$\alpha=90^{\circ}$ (transport over the poles) are discussed.
The length of the integration is 12 days which corresponds to one complete revolution 
of the cosine bell around the sphere.  The base resolution for both shallow water models 
is a $5^{\circ} \times 5^{\circ}$ coarse mesh with additional initial refinements surrounding 
the cosine bell. In particular, the base resolution in SEM is represented by
$54$ elements (see Table~\ref{tab:sem_resolution}) whereas the adaptations in FV start 
from $6 \times 8$ initial blocks.
We test the refinement levels 1, 2, 3 and 4 which are equivalent to the 
resolutions $2.5^{\circ} \times 2.5^{\circ}$, $1.25^{\circ} \times 1.25^{\circ}$, $0.625^{\circ} 
\times 0.625^{\circ}$  and $0.3125^{\circ} \times 0.3125^{\circ}$ within the refined areas. 
Note that these grid spacings only represent approximate resolutions for SEM because of the clustering of the 
collocation points near the element boundaries (see Fig.~\ref{fig:sem_gridpoints}). 

In both models  we refine the grid if the geopotential height field 
of the cosine bell exceeds $h \ge 53$ m. This value corresponds to approximately 5\% of 
the initial peak value with $h_{max} = 1000$ m. Of course, other refinement thresholds are also feasible.
The threshold has been chosen since the 
refined area now tightly surrounds the cosine bell in the regular latitude-longitude grid 
(model FV). In SEM though, the adaptations are padded by a one-element wide halo.
 This leads to a broader refined area in comparison to FV but less evaluations of the 
 refinement criterion. For the FV model, the refinement criterion is evaluated at every time step. 
 The grid is refined up to the maximum refinement level whenever the criterion is met. Grid 
 coarsenings are invoked after the cosine bell left the refined area and consequently, the criterion 
 is no longer fulfilled. The time steps in the FV simulations are variable and match a CFL 
 stability criterion of  $|$CFL$| \le 0.95$ at the finest refinement level. This setup corresponds exactly 
 to the adaptive advection experiments with FV  in \cite{Jablonowski:2006}. For the SEM model, 
 the time-step is constant and restricted by the CFL number computed on the finest grid. 

\begin{figure}
  \centering
  \includegraphics[width=1.0\textwidth]{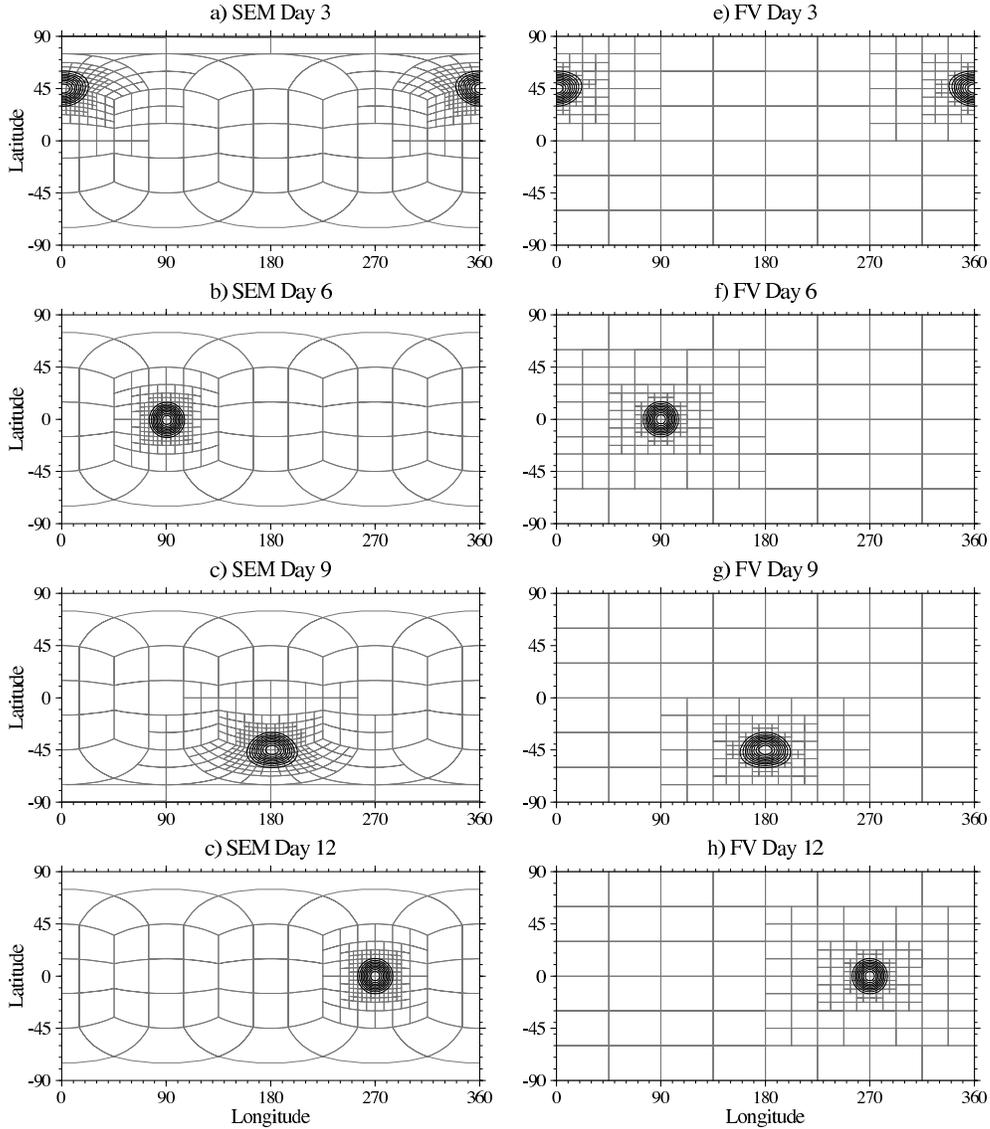}
  \caption{Snapshots of the cosine bell (test case 1) with rotation angle $\alpha=45^{\circ}$ 
  and three refinement levels at day 3, 6, 9 and 12. The finest grid spacing is $0.625^{\circ} 
  \times 0.625^{\circ}$, the adapted spectral elements (SEM) and blocks (FV) are overlaid. 
  Left column (a-d) shows the SEM model, right column (e-h) depicts the FV model. The 
  contour interval is 100 m, the zero contour is omitted.}
  \label{fig:sw_test1}
\end{figure}

Figure~\ref{fig:sw_test1} shows snapshots of the cosine bell with rotation angle $\alpha = 45^{\circ}$  and 
three refinement levels ($0.625^{\circ} \times 0.625^{\circ}$) at day 3, 6, 9 and 12. The SEM 
model with its refined spectral elements on the cubed sphere is depicted in the left column 
(Fig.~\ref{fig:sw_test1}a-d), the right column (Fig.~\ref{fig:sw_test1}e-g) displays the FV model 
with its block-structured latitude-longitude grid. Note that each spectral element (SEM) 
contains additional $6 \times 6$ GL grid points, whereas each block (FV) consists of $6 \times 9$ 
grid points in the latitudinal and longitudinal direction, respectively. 
In both models the initial state (not shown) resembles the final state (day 12) very closely. It 
can clearly be seen that the adapted grids of both models reliably track the the cosine bell 
without visible distortion or noise.

\begin{figure}
  \centering
  \includegraphics{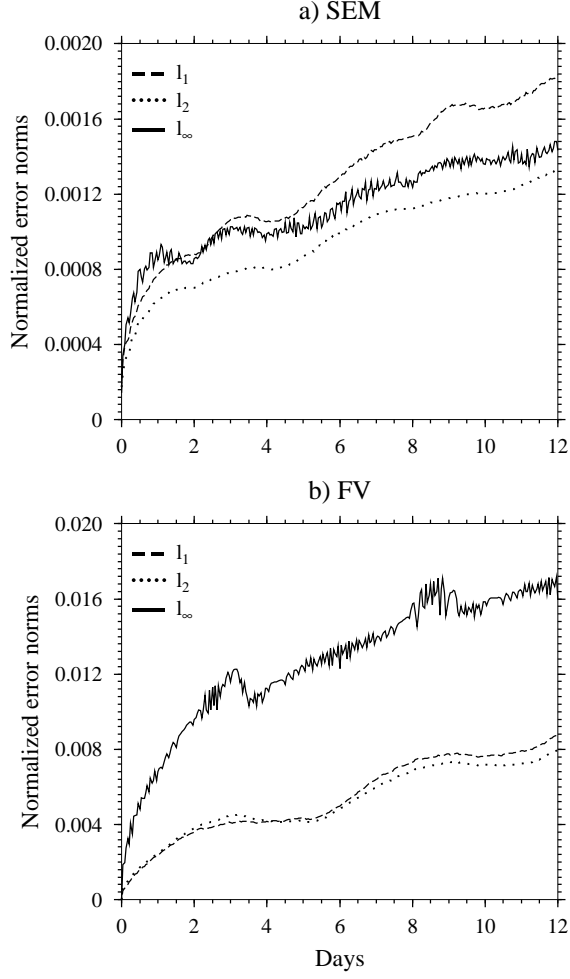}
  \caption{Time traces of the normalized $l_1$, $l_2$ and $l_{\infty}$ geopotential height 
  error norms for the cosine bell advection test (test case 1) with $\alpha = 45^{\circ}$. The (a) 
  SEM and (b) FV advection models utilized three refinement levels that correspond to the 
  finest grid resolution $0.625^{\circ} \times 0.625^{\circ}$. Note that the scales in (a) and (b) 
  differ by a factor of 10.}
  \label{fig:sw_test1_norms}
\end{figure}

The errors can be further assessed in Fig.~\ref{fig:sw_test1_norms} which shows the time 
traces of the normalized $l_1$, $l_2$ and $l_{\infty}$ norms of the geopotential height field 
for rotation angle $\alpha = 45^{\circ}$. At this rotation angle the errors in SEM are an order 
of magnitude smaller than the errors in FV. This is mainly due to the operator-splitting 
approach in FV which produces the maximum error at $\alpha=45^{\circ}$. It is apparent that
SEM transports the cosine bell rather smoothly across the corners of the cubed mesh.
The errors in FV are considerably lower for rotation angle $\alpha=0^{\circ}$. 
This is shown in Table~\ref{table:swtc1:sem:fv} that 
documents the $\alpha=0^{\circ}$ normalized height errors as well as the absolute maximum 
and minimum of the height field after one full revolution for the refinement levels 1 through 4. The errors 
of both models at any refinement level are now comparable to each other. In addition, the 
table shows that SEM introduces undershoots during the simulation which are documented 
by the negative minimum height values $h_{min}$. This is not the case for the monotonic 
and conservative FV advection algorithm. Here it can also be seen that the maximum height 
of the cosine bell $h_{max}$ in FV is affected by the monotonicity constraint which acts as 
a nonlinear dissipation mechanism. It consequently reduces the peak amplitude of the 
cosine bell, especially at lower resolutions when the bell is not well-resolved. This decrease 
in maximum amplitude is less profound in SEM. 

\begin{table}[htdp]
\caption{Normalized height error norms and characteristics for the advection of a cosine 
bell (test case 1) after one revolution (day 12) at different refinement levels and rotation 
angle $\alpha=0^{\circ}$ . The first column indicates the resolution of the grid within the 
refined area. Both the SEM model (top) and FV model (bottom) are shown.}
\begin{center}
\begin{tabular}{lcccc}
\hline \hline
Resolution      &$l_1$              &$l_2$              &$l_\infty$ &$h(m)$ $\max$/$\min$ \\ \hline
 \multicolumn{5}{l}{SEM} \\
$2.5^{\circ} \times 2.5^{\circ}$		&0.0503   	&0.0269   &0.0195 & 991.6/-15.1\\
$1.25^{\circ}  \times  1.25^{\circ}$       	&0.0085   &0.0056   &0.0057 & 997.5/-4.2\\
$0.625^{\circ} \times  0.625^{\circ}$       &0.0019   &0.0014   &0.0019 & 999.1/-1.1\\
$0.3125^{\circ} \times  0.3125^{\circ}$     &0.0008   &0.0006   &0.0015 & 999.7/-0.9 \\ \hline
\multicolumn{5}{l}{FV} \\
$2.5^{\circ} \times 2.5^{\circ}$		&0.0341 &0.0301 &0.0317 &949.1/0 \\
$1.25^{\circ} \times 1.25^{\circ}$		&0.0097 &0.0103 &0.0150 &984.2/0 \\
$0.625^{\circ} \times 0.625^{\circ}$	&0.0016 &0.0021 &0.0044 &995.0/0 \\
$0.3125^{\circ} \times 0.3125^{\circ}$	&0.0003 &0.0005 &0.0014 &998.4/0  \\ \hline \hline
\end{tabular}
\label{table:swtc1:sem:fv}
\end{center}
\end{table} 

\begin{figure}
  \centering
  \includegraphics{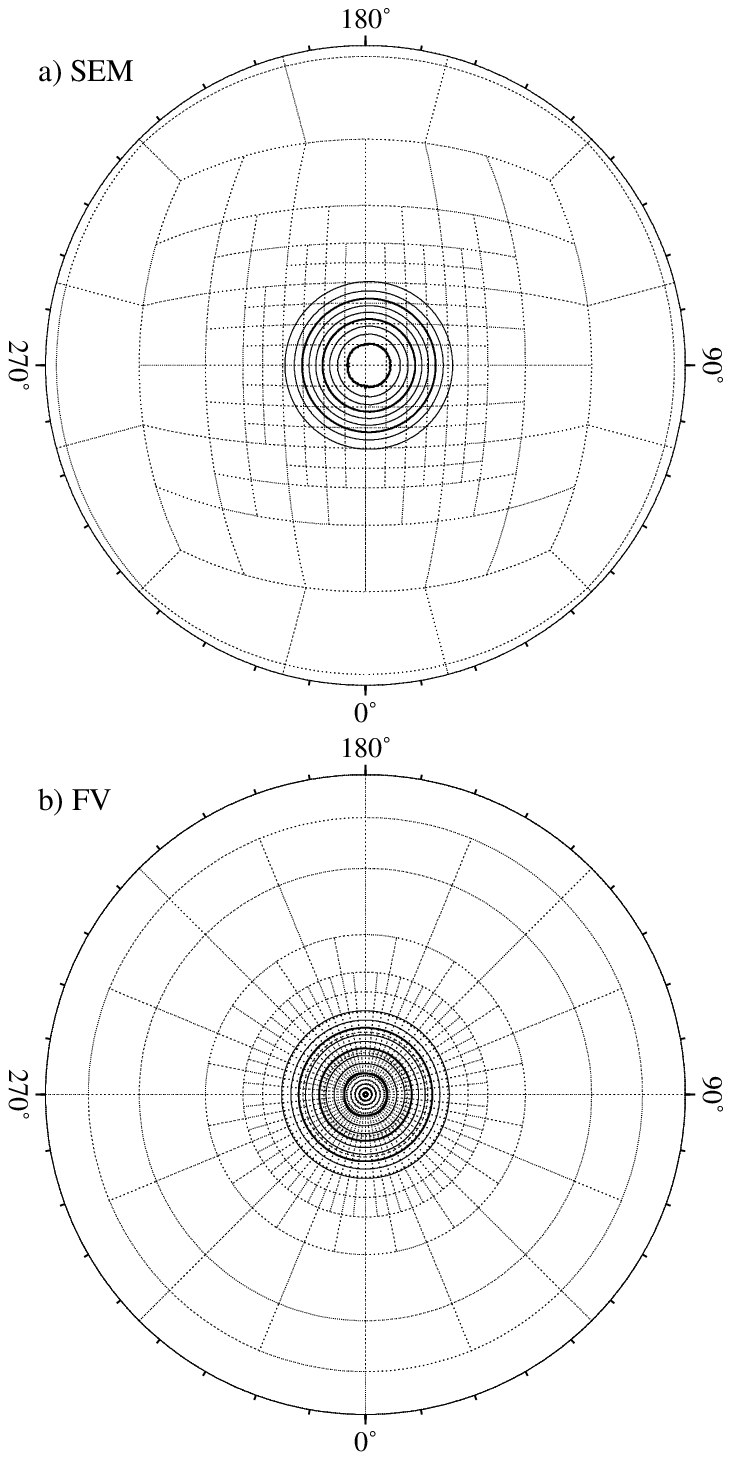}
  \caption{North-polar stereographic projection of the cosine bell (test case 1) with rotation angle $\alpha=90^{\circ}$ 
  and three refinement levels at day 3 for (a) SEM and (b) FV. The finest grid spacing is $0.625^{\circ} 
  \times 0.625^{\circ}$, the adapted spectral elements (SEM) and blocks (FV) are dotted. 
  The contour interval is 100 m, the zero contour is omitted.}
  \label{fig:sw_test1_90}
\end{figure}

Another snapshot of the cosine bell advection test at day 3 with three refinement levels
is shown in Fig.~\ref{fig:sw_test1_90}. Here, the
rotation angle is set to $\alpha=90^{\circ}$ which directs the bell straight over the poles. At day 3, the bell
reaches the North Pole and both models refine the grid reliably over this region. Nevertheless, a distinct difference 
between the two models is the total number of refined elements (SEM) and blocks (FV) needed at high latitudes.
While SEM (Fig. ~\ref{fig:sw_test1_90}a) keeps the total number of adapted elements small in the chosen cubed-sphere
geometry, the FV model (Fig. ~\ref{fig:sw_test1_90}b)
needs to refine a large number of blocks due to the convergence of the meridians in the latitude-longitude mesh.
This leads to an increase in the computational workload in the adaptive FV experiment due to the increased number of
blocks and small time steps needed in polar regions. In this FV run the time step varies between approximately 
$\Delta t = 600$ s and $\Delta t = 10$ s which strongly depends on the position of the cosine bell \citep{Jablonowski:2006}.
The SEM model though utilizes a constant time step of $\Delta t = 10$ s. Further statistics for
test case 1 are listed in Table~\ref{tab:statistics}. In particular, for rotation angle $\alpha=45^{\circ}$ the 
number of refined blocks in FV is smaller than the number of refined elements in SEM. This effect is
mainly due to the additional halo region in SEM.

\subsection{Steady-state geostrophic flow}
Test case $2$ is a steady-state zonal geostrophic flow, representing a balance 
between the Coriolis and geopotential gradient forces in the momentum equation.  
The initial velocity field on the sphere is given by
\begin{align*}
u  &=  u_0 \left(\: \cos\theta \cos\alpha + 
\cos \lambda\sin\theta\sin\alpha \:\right) \\
v  &=  - u_0 \sin\lambda \sin\alpha~.
\end{align*}
where $\alpha$ is the angle between the axis of solid
body rotation and the polar axis. The analytic geopotential
field $\Phi = gh$ is specified as
\[
\Phi = \Phi_0 - 
\left(\: a \Omega u_0 + \frac{u_0^2}{2}\:\right) \times
\left(\:-\cos\lambda\cos\theta\sin\alpha + \sin\theta\cos\alpha\:\right)^2~.
\]
$a$ is the radius of the earth and
$\Omega$ is the rotation rate. Parameter values are 
specified to be $u_0 = 2\pi a/(12\:{\rm days})$ and 
$\Phi_0 = 2.94\times 10^4$ ${\rm m}^2/{\rm s}^2$. These fields also represent the
analytic solution of the flow. The
Coriolis parameter is given by
\[
f = 2\Omega\left(\:
-\cos\lambda\cos\theta\sin\alpha + \sin\theta\cos\alpha\:
\:\right).
\]

We primarily use this large-scale flow pattern to assess the characteristics of the fine-coarse 
mesh interfaces in two statically adapted grid configurations. Both models SEM and FV 
utilize non-conforming meshes with a resolution jump by a factor of two in each direction at 
the interfaces. Note that the FV model requires interpolations in the ghost cell regions of neighboring 
blocks whenever the resolution is changed. This is not the case for SEM. In general, inserting 
a refined patch of elements (SEM) or blocks (FV) in a random location should result in either 
no changes in the error, if the flow is completely resolved, or in a small decrease in the overall 
error. Inserting the patch in a strategic location might even lower the error more significantly.

Here, we start our simulations from a uniform $2.5^{\circ} \times 2.5^{\circ}$ base grid which 
is given by $216$ spectral elements with $6\times 6$ GL points in SEM. In FV, 
this corresponds to the block data structure consisting of $12 \times 16 = 192$ blocks with 
$6 \times 9$ grid points in latitudinal and longitudinal direction, respectively. Two 
refinement levels are utilized which lead to the finest mesh spacing 
$0.625^{\circ} \times 0.625^{\circ}$ in both models.

\begin{figure}
  \centering
  \includegraphics{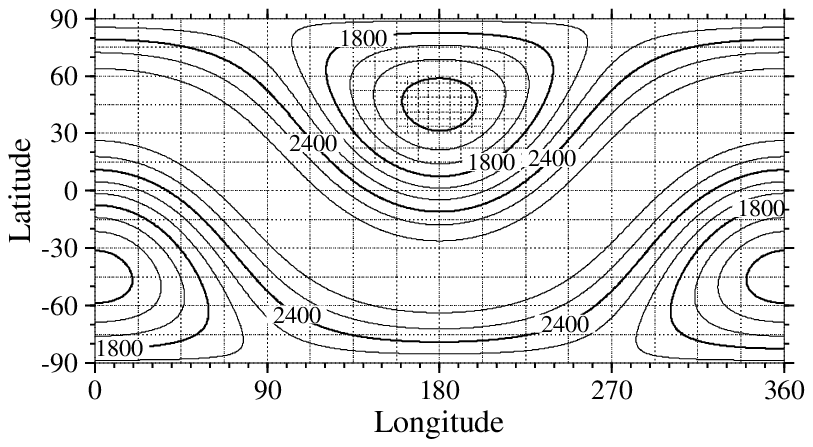}
  \caption{Geopotential height field for test case 2 with $\alpha=45^{\circ}$ at day 14 as simulated with the FV model. The statically
  adapted blocks with 2 refinement levels are overlaid (dotted contour). The refined patch is 
  centered at (180$^{\circ}$E, 45$^{\circ}$N), the finest grid spacing is $0.625^{\circ}  \times 0.625^{\circ}$.
  The contour interval is 200 m.}
  \label{fig:sw_test2_h}
\end{figure}

In the first configuration, a refined patch of size $45^{\circ} \times 30^{\circ}$ (longitudes 
$\times$ latitudes) is centered at (180$^{\circ}$E, 45$^{\circ}$N). This patch spans the 
domain (157.5$^{\circ}$E,30$^{\circ}$N)-(202.5$^{\circ}$E,60$^{\circ}$N) which is shown by
the dotted contours of the adapted FV blocks in Fig.~\ref{fig:sw_test2_h}. Here, the geopotential height 
field of the FV simulation at day 14 is displayed which is visually indistinguishable from the initial state.
In the second configuration a patch of identical size is centered at (135$^{\circ}$E, 30$^{\circ}$N). This 
covers the region between (112.5$^{\circ}$E,15$^{\circ}$N) - (157.5$^{\circ}$E,45$^{\circ}$N)
which is characterized by strong gradients in the geopotential height field (see also the discussion in
\cite{Jablonowski:2004}). The 
two locations give insight into the dependency of the errors on the position of the refined 
patch.

\begin{figure}
  \centering
  \includegraphics{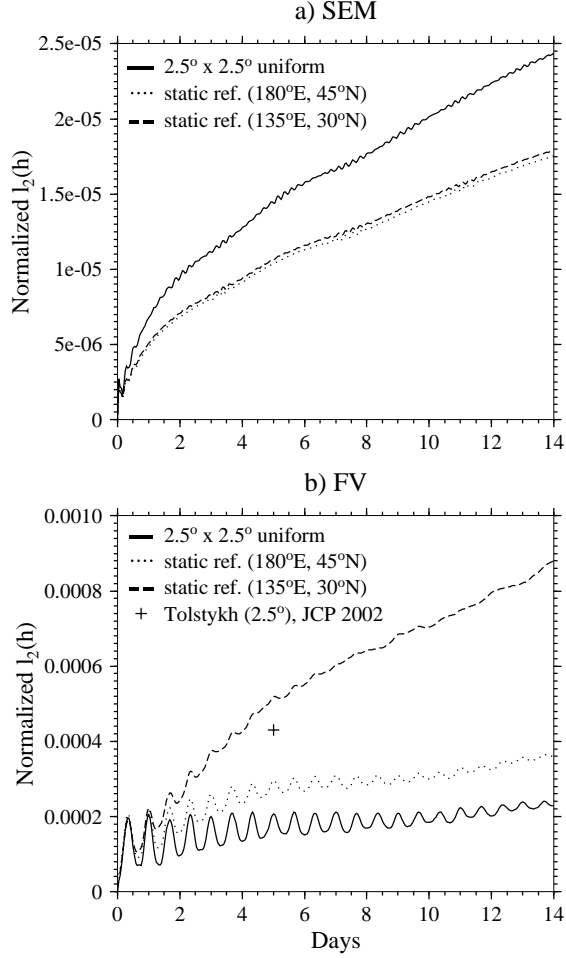}
  \caption{Time traces of the normalized $l_2$ geopotential height error norms for the steady 
  state geostrophic flow (test case 2) with rotation angle  $\alpha = 45^{\circ}$. Two 
  simulations with static refinement regions centered at (180$^{\circ}$E, 45$^{\circ}$) and 
  (135$^{\circ}$E, 30$^{\circ}$) are compared to a uniform $2.5^{\circ} \times 2.5^{\circ}$ 
  resolution run. The adaptive (a) SEM and (b) FV model runs start with a 
  $2.5^{\circ} \times 2.5^{\circ}$ base grid with two refinement level (finest grid resolution 
  is $0.625^{\circ} \times 0.625^{\circ}$.)} \label{Fig:swtc2_norms}
\end{figure}

Both statically adapted configurations are integrated for $14$ days with $\alpha=45^{\circ}$. 
As mentioned before this rotation angle represents the most challenging direction for both models due to the choice 
of the cubed-sphere geometry in SEM and the operator-splitting approach in FV. The results 
for the $l_2$ normalized geopotential height errors for SEM and FV are reported in 
Fig.~\ref{Fig:swtc2_norms}. It is expected that the errors in SEM are lower than in FV which 
is confirmed by Figs.~\ref{Fig:swtc2_norms}a-b. This is due to the spectral convergence (SEM) 
of the smooth solution which is infinitely differentiable. In general, such spectral convergence 
can not be archived by grid point models which typically exhibit higher error norms. Here it 
is interesting to note that the two refined patches in SEM lead to a slight decrease in the error 
in comparison to the uniform-mesh run (Fig.~\ref{Fig:swtc2_norms}a). The errors are 
independent of the location of the refined area. This is in contrast to the error norms in FV 
(Fig.~\ref{Fig:swtc2_norms}b). The refined patches introduce slight disturbances of the 
geostrophic balance and non-divergent wind field which now cause the error norms to increase 
in comparison to the uniform-mesh run. The increase is sensitive to the location of the 
refined patch. In particular the errors in FV grow faster if the patch intersects the strong 
gradient regime in the geopotential height field (centered at (135$^{\circ}$E, 30$^{\circ}$N)). 
Despite this characteristic, the error is still in the expected range for grid point based models, 
as for example compared to \cite{tolstykh:02}. The increase in error is mainly triggered 
by interpolations in the ghost cell regions (see also \cite{Jablonowski:2004fa,Jablonowski:2006b}) 
and will be further investigated.

\subsection{Flow over an idealized mountain}
\label{subsec:test5}
\begin{figure}
  \centering
  \includegraphics[width=1.0\textwidth]{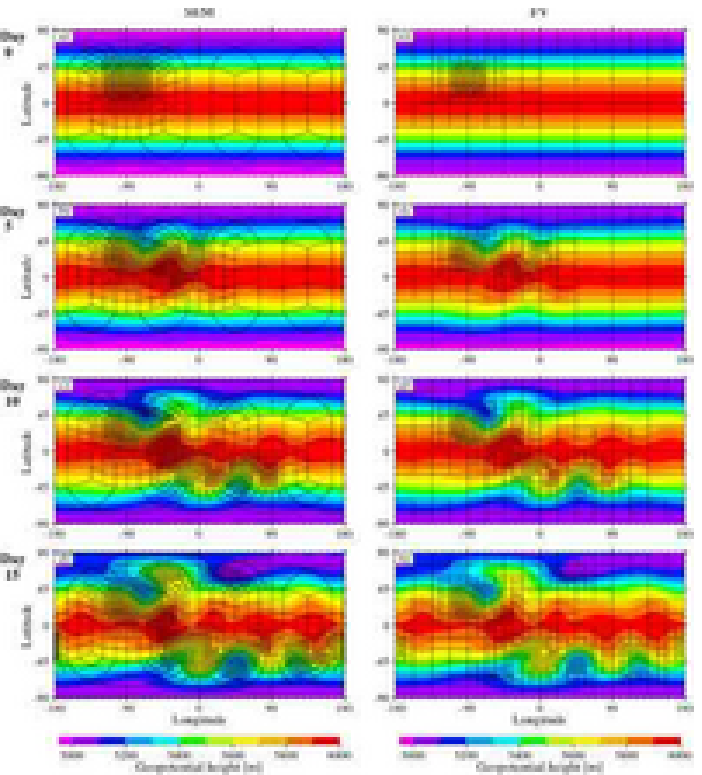}
  \caption{Snapshots of the geopotential height field (test case 5, flow over a mountain) with 
  three refinement levels at day 0, 5, 10 and 15. The finest grid spacing is 
  $0.625^{\circ} \times 0.625^{\circ}$, the adapted spectral elements (SEM) and blocks (FV) 
  are overlaid. Left column (a-d) shows the SEM model, right column (e-h) depicts the FV 
  model. The refinement criterion is $|\zeta| \ge 2 \times 10^{-5}$ s$^{-1}$. The contour interval 
  is $100$ m.}\label{fig:sw_test5}
\end{figure}

Test case 5 of the \cite{Williamson:1992ts} test suite is a zonal flow impinging on a 
mountain. The mean equivalent depth of the atmosphere is set to $h_0 = 5960$ m. 
The mountain height is given by $h_s = h_{s_0}(1 - r/R)$, where 
$h_{s_0} = 2000$ m, $R = \pi/9$, and 
$r^2 = \min[R^2,(\lambda-\lambda_c)^2 + (\theta - \theta_c)^2]$. The 
center of the mountain is located at $\lambda_c = 3\pi/2$ and 
$\theta_c = \pi/6$ in spherical coordinates.  The test case is integrated for 15 model days 
with a $5^{\circ} \times 5^{\circ}$  base grid as in test case 1. 
In addition, three levels of static 
refinements are introduced where the height of the mountain is  greater than $0$ m. 
The corresponding initial grids for SEM and FV are shown in 
Figs.~\ref{fig:sw_test5}a, \ref{fig:sw_test5}e and \ref{fig:amr_grids}.

During the course of the simulation dynamic refinements with three refinement levels are 
applied whenever the absolute value of the relative vorticity $\zeta$ is greater than  
$2 \times 10^{-5}$ s$^{-1}$. This refinement criterion is evaluated every two hours.
Grid coarsenings, on the other hand, are invoked in regions where the threshold 
is no longer met. Figure \ref{fig:sw_test5} shows the time evolution of the geopotential height 
field for both models at days 0, 5, 10 and 15. The adapted elements on the cubed sphere 
(SEM) and adapted blocks (FV) are overlaid. It can be seen that both models track the 
evolving lee-side wave train reliably and place most of the adaptations in the Southern 
Hemisphere at day 15. Some differences in the refinement regions are apparent. Overall, 
the refinements in SEM cover a broader area because a one-element wide halo was 
enforced around the regions marked for refinement. This option reduces the number 
of adaptation cycles but on the other hand increases the overall workload. 
The number of adapted elements in SEM and adapted blocks in FV at day 15
is also documented in Table~\ref{tab:statistics}. They almost differ by a factor of 2.
Note, that SEM employs a $20$ s time step, the time step for FV is $138$ s.

\begin{figure}
  \centering
  \includegraphics{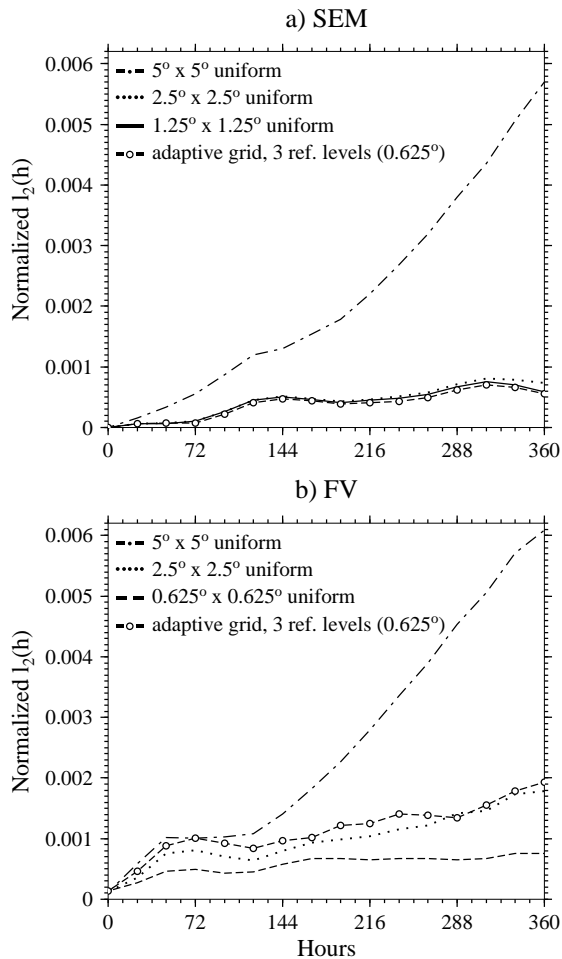}
  \caption{Time traces of the normalized $l_2$ geopotential height error norms for the flow over 
  a mountain (test case 5). The adaptive simulations with three refinement levels 
  ($0.625^{\circ} \times 0.625^{\circ}$ at the finest level) is compared to uniform resolution 
  runs.}\label{Fig:swtc5_norms}
\end{figure}

The results can be quantitatively compared via normalized error norms which are shown 
in Fig.~\ref{Fig:swtc5_norms}. An analytic solution is not  known. Therefore, the normalized 
error metrics are computed by comparing the SEM and FV simulations to a T$213$ 
spectral transform reference solution provided by the National Center for 
Atmospheric Research (NCAR) \citep{jakob:93,jakob:95}. The latter 
is available as an archived netCDF data set with daily snapshots of the spectral transform 
simulation. The T$213$ spectral simulation utilized a Gaussian grid with $320 \times 640$ 
grid points in latitudinal and longitudinal direction which corresponds to a grid spacing of 
about 55 km at the equator. Figure~\ref{Fig:swtc5_norms} compares the normalized 
$l_2$ height errors of the adaptive runs to uniform-resolution simulations. Both models are 
depicted. It is interesting to note that the errors in SEM (Fig.~\ref{Fig:swtc5_norms}a) already 
converge within the uncertainty of the NCAR reference solution (see \cite{Taylor:1997vn}) at 
the uniform resolution $2.5^{\circ} \times 2.5^{\circ}$. The SEM adaptive run matches this 
error trace very closely. On the other hand, the errors in FV (Fig.~\ref{Fig:swtc5_norms} b)
converge within the uncertainty of the reference solution at the finer uniform resolution 
$0.625^{\circ} \times 0.625^{\circ}$. Here, the error trace of the FV adaptive simulation 
resembles the $2.5^{\circ} \times 2.5^{\circ}$ uniform run despite the higher resolution in 
the refined areas. It indicates that the coarser domains in FV still contribute considerably to 
the global $l_2$ error measure. In addition, the interpolations in the ghost cell regions add to 
the error norms.

\subsection{Rossby-Haurwitz wave}
The initial condition for test case 6 of the \cite{Williamson:1992ts} test suite is a wavenumber 
4 Rossby-Haurwitz wave. These waves are analytic solutions to the nonlinear 
non-divergent barotropic vorticity equation, but not closed-form solutions of the barotropic 
shallow water equations. However, in a shallow water system the wave pattern moves 
from west to east without change of shape during the course of the integration. The initial 
conditions are fully described in \cite{Williamson:1992ts}, the orography field is set to zero. 
The Rossby-Haurwitz wave with wavenumber 4 exhibits extremely strong gradients in both 
the geopotential and the wind fields. This test is especially hard for the FV adaptive grid 
simulations due not only to the strong gradients but also to the dominant $45^{\circ}$ 
transport angle of the flow in midlatitudes. This challenges the underlying 
operator splitting approach in FV and accentuates even minor errors.

Both shallow water models are integrated for 14 days on a $2.5^{\circ} \times 2.5^{\circ}$ base grid.
This base resolution is identical to test case 2. 
In addition, static refinements at refinement 
level 1 $(1.25^{\circ} \times 1.25^{\circ})$ are placed within eight pre-determined regions 
of interest. In particular, the grid is refined where the initial meridional wind field is $v \le -60$ 
m s$^{-1}$ which leads to an almost identical number of refined elements and blocks in the
two models (Table~\ref{tab:statistics}).
This rather arbitrary refinement criterion is intended to test whether the 
wavenumber 4 pattern moves smoothly through the refined grid patches. As an aside, 
minor improvements of the error norms are expected. The daily simulation results of SEM and 
FV are compared to the NCAR reference solution \citep{jakob:93,jakob:95}. As for test case 
5 (flow over the mountain), the latter is available as an archived netCDF data set that contains 
the daily results of an NCAR spectral shallow water model at the resolution T$213$ 
($\approx$ 55 km). 

\begin{figure}
  \centering
  \includegraphics{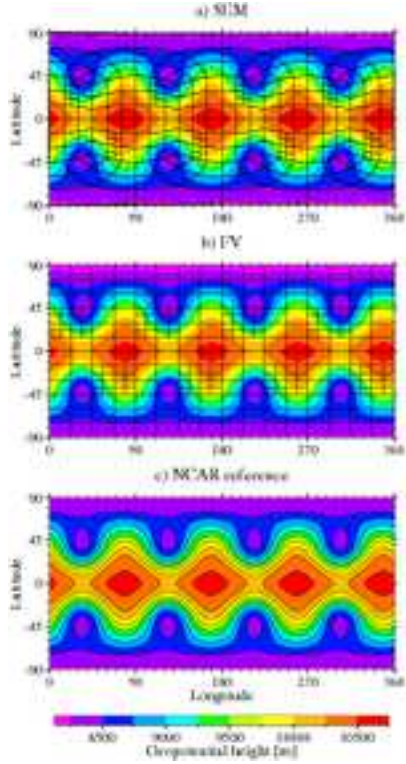}
  \caption{Snapshots of the geopotential height field of the Rossby-Haurwitz wave (test case 6) 
  at day 7 for (a) SEM, (b) FV and (c) the NCAR spectral model (reference solution). SEM and 
  FV utilize an adaptive grid with one level of static adaptations. The base resolution is 
  $2.5^{\circ} \times 2.5^{\circ}$ and $1.25^{\circ} \times 1.25^{\circ}$ within the refined 
  region. The adapted elements (SEM) and blocks (FV) are overlaid. The contour interval is 
  $250$ m.}\label{fig:sw_test6}
\end{figure}

Figures~\ref{fig:sw_test6}a-b show the geopotential height field of the adaptive SEM and 
FV model runs at day 7. It can clearly be seen that both SEM and FV maintain the wavenumber 
4 pattern of the height field rather well while moving  through the refined patches. Here, the 
results can be visually compared to the NCAR reference solution (Fig.~\ref{fig:sw_test6}c). 
No noise or distortions are visible at this stage. However, FV develops slight asymmetries in 
the height field at later times. They originate at the interfaces of the refined areas and 
slightly disturb the wave field over the course of the integration. These  perturbations can again 
be traced back to ghost cell interpolations in the FV model
that now lead to a slight increase in the $l_2$ errors norm. The effect is amplified by the 
very strong gradients in the flow field and the dominant $45^{\circ}$ transport angle. The time
steps for these simulations are $15$ s and $225$ s for SEM and FV, respectively.

\begin{figure}
  \centering
  \includegraphics{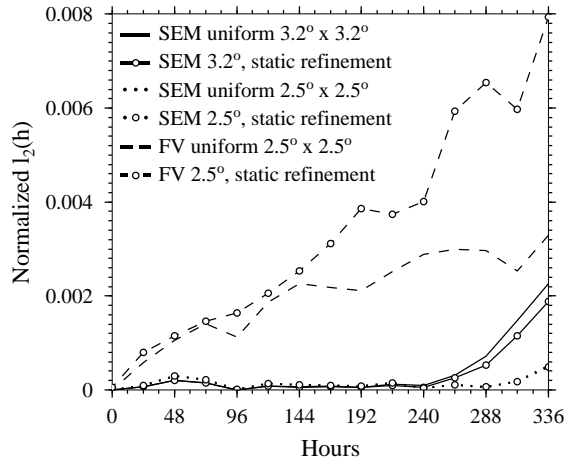}
  \caption{Time traces of the normalized $l_2$ geopotential height error norms for the 
  Rossby-Haurwitz wave (test case 6). The statically adaptive simulations with one 
  refinement levels ($1.25^{\circ} \times 1.25^{\circ}$ at the finest level) are compared to 
  SEM and FV uniform resolution runs.}
  \label{Fig:swtc6_norms}
\end{figure}

The error norms for both models are quantitatively compared in Fig.~\ref{Fig:swtc6_norms}. 
The figure shows the time evolution of the normalized $l_2$ height error norms in comparison to 
fixed-resolution SEM and FV model runs. In particular, two different base resolutions are shown for SEM.
These are the regular $2.5^{\circ} \times 2.5^{\circ}$ base grid (216 elements) and a coarser $3.2^{\circ} \times 3.2^{\circ}$
base mesh (96 elements). In both SEM cases, $6 \times 6$ GL quadrature points
per spectral element are used. The coarser simulation is added since SEM already converges within
the uncertainty of the reference solution \citep{Taylor:1997vn} 
on the $2.5^{\circ}$ base grid.
Figure~\ref{Fig:swtc6_norms} shows that SEM exhibits smaller errors than FV at the same uniform resolution. At day
12, the errors in the FV run are rather similar to the coarser $3.2^{\circ}$ SEM simulation. 
With respect to the static adaptations, SEM does not show an increase in the error 
in comparison to FV. Instead, the SEM errors in the statically adaptive 
runs are almost identical to the uniform-mesh simulation or slightly diminish at the end of 
the forecast period. 

\section{Conclusions}
\label{sec:conclusion}
In this paper, two shallow water models with adaptive mesh refinement capabilities
are compared. The models are an interpolation-based spectral 
element model (SEM) on a cubed-sphere grid and a conservative and monotonic 
finite volume model (FV) in latitude-longitude geometry. Both adaptive mesh approaches 
utilize a quad-tree AMR technique that reduces the mesh spacings by a factor of two during each refinement step. 
Coarsenings reverse this adaptations principle. Then four ``children/leaves'' are coalesced
which doubles the grid distances. 
In SEM, the refinement strategy  targets the spectral elements which contain additional
Gauss-Legendre and Gauss-Lobatto-Legendre collocation points for scalar and vector components.
In FV, the adaptations are applied to a block-data structure in
spherical coordinates which consists of a fixed (self-similar) block size. These blocks
are surrounded by ghost cell regions which require interpolation and averaging procedures at fine-coarse
mesh interfaces. No ghost cells areas are needed in SEM.
In both models neighboring elements or blocks can only differ by one refinement level.
This leads to cascading and non-conforming refinement regions around the finest mesh.

The models are compared via selected test cases from the standard \cite{Williamson:1992ts}  shallow water test suite. 
They include the advection of a cosine bell (test 1), a steady-state 
geostrophic flow (test 2), a flow over a mountain (test 5) and a Rossby-Haurwitz wave (test 6). Both static and dynamics adaptations
are assessed which reveal the
strengths and weaknesses of the AMR approaches.
The AMR simulations show that both models successfully place static and dynamic adaptations in
local regions without requiring a fine grid in the global domain. 
The adaptive grids  reliably track the user-selected features of interests without visible distortions or noise
at the mesh interfaces. In particular, 
two dynamic refinement criteria were evaluated. Among them were a simple
geopotential height threshold  (test 1) as well as the magnitude of the
relative vorticity  (test 5). The latter successfully steered the SEM and FV refined grids into the 
Southern Hemisphere at the end of the 15-day forecast period. In addition, two static AMR configurations in user-determined 
regions of interest were tested. They confirmed that the flows move smoothly through the refined areas in both SEM and FV. Nevertheless,
the FV simulations showed that small errors originate at fine-coarse grid interfaces. These errors were
due to the interpolation and averaging mechanisms in the ghost cell regions of the block-data structure. 
This was not the case in SEM which exhibited a decrease in error whenever adaptations were introduced.

Overall, the number of dynamically refined elements or blocks was comparable in both models if the adaptations
were confined to the equatorial or midlatitudinal regions. In polar regions though, the number of refined blocks
in FV exceeded the number of refined elements in SEM considerably due to the convergence of the meridians in the
latitude-longitude grid. 
In general, it was shown that SEM exhibited smaller errors than FV for all test cases at identical resolutions.
This is expected for SEM's high-order numerical technique despite its non-conservative and non-monotonic 
nature. The latter can cause spurious oscillations and negative values for positive-definite fields (test 1).
In contrast, the FV technique is mass-conservative and monotonic which, on the other hand, 
introduces numerical dissipation through its monotonicity constraint. The dissipation is reduced  at high resolutions.
Then the SEM and FV simulations resemble each other very closely.

The problem of conservation and the removal of some of the oscillations in SEM can be addressed 
by using the discontinuous Galerkin (DG) formulation. 
Future developments will include an evaluation of the DG method as well as efficient time-stepping techniques 
\citep{Nair:2005aj,St-Cyr:2005jw,St-Cyr:2005ur, St-Cyr:2006xf} to avoid the very small time steps for numerical
stability. In addition, tests involving strong 
jets such as in \cite{Galewsky:2004wa} and \cite{laeuter:05}, effects of orography and a 
full GCM comparison will be the subject of interesting future research.
\newpage
\section*{Acknowledgments.}
The authors thanks Catherine Mavriplis and Paul Fischer for many insightful 
discussions on adaptive spectral elements methods. This work was partially 
supported by an NSF collaborations in mathematics and the geosciences grant 
(022282) and the DOE climate change prediction program (CCPP). In addition,
CJ was partly supported by NASA Headquarters under the Earth System Science Fellowship
Grant NGT5-30359.

\bibliography{bibamr}

\begin{thebibliography}{10}

\bibitem{bacon:2000}
{\sc Bacon, D.~P., Ahmad, N.~N., Boybeyi, Z., Dunn, T.~J., Hall, M.~S., Lee, P.
  C.~S., Sarma, R.~A., Turner, M.~D., III, K. T.~W., Young, S.~H., and Zack,
  J.~W.}
\newblock A dynamically adapting weather and dispersion model: The operational
  multiscale environment model with grid adaptivity ({OMEGA}).
\newblock {\em Mon.Wea. Rev. 128\/} (2000), 2044--2076.

\bibitem{behrens:2006}
{\sc Behrens, J.}
\newblock {\em Adaptive Atmospheric Modeling - {K}ey techniques in grid
  generation, data structures, and numerical operations with applications}.
\newblock Springer, 2006.
\newblock {L}ecture Notes in Computational Science and Engineering, ISBN
  3-540-33382-7, 207 pp.

\bibitem{behrens:2005}
{\sc Behrens, J., Rakowsky, N., Hiller, W., Handorf, D., L{\"a}uter, M.,
  P{\"a}pke, J., and Dethloff, K.}
\newblock amatos: {P}arallel adaptive mesh generator for atmospheric and
  oceanic simulation.
\newblock {\em Ocean Modelling 10\/} (2005), 171--183.

\bibitem{Boyd:1996hn}
{\sc Boyd, J.~P.}
\newblock The {E}rfc-{L}og filter and the asymptotics of the {E}uler and
  {V}andeven sequence accelerations.
\newblock In {\em Proceedings of the Third International Conference on Spectral
  and High Order Methods\/} (1996), A.~V. Ilin and L.~R. Scott, Eds., Third
  International Conference on Spectral and High Order Methods, Houston Journal
  of Mathematics, pp.~267--276.

\bibitem{Boyd:2004lc}
{\sc Boyd, J.~P.}
\newblock Book review: {H}igh-order methods for incompressible fluid flow.
  \newblock.
\newblock {\em SIAM Rev. 46}, 1 (2004), 151--157.

\bibitem{carpenter:90}
{\sc Carpenter, R.~L., Droegemeier, K.~K., Woodward, P.~R., and Hane, C.~E.}
\newblock Application of the {P}iecewise {P}arabolic {M}ethod ({PPM}) to
  meteorological modeling.
\newblock {\em Mon. Wea. Rev. 118\/} (March 1990), 586--612.

\bibitem{Colella:1984hi}
{\sc Colella, P., and Woodward, P.~R.}
\newblock The piecewise parabolic method ({PPM}) for gas-dynamical simulations.
\newblock {\em J. of Comput. Phys. 54\/} (1984), 174--201.

\bibitem{dennis:03}
{\sc Dennis, J.~M.}
\newblock Partitioning with space-filling curves on the cubed sphere.
\newblock In {\em Proc. International Parallel and Distributed Processing
  Symposium (IPDPS)\/} (Nice, France, April 2003), IEEE/ACM.
\newblock {CD-ROM}. No. 269a.

\bibitem{Deville:2002fn}
{\sc Deville, M.~O., Fischer, P.~F., and Mund, E.~H.}
\newblock {\em High-{O}rder {M}ethods for {I}ncompressible {F}luid {F}low}.
\newblock Cambridge monographs on applied and computational mathematics.
  Cambridge University Press, 2002.

\bibitem{Dietachmayer:1992tg}
{\sc Dietachmayer, G.~S., and Droegemeier, K.~K.}
\newblock Application of continuous dynamic grid adaptation techniques to
  meteorological modeling. part i: Basic formulation and accuracy.
\newblock {\em Mon. Wea. Rev. 120\/} (1992), 1675--1706.

\bibitem{Feng:2002uq}
{\sc Feng, H., and Mavriplis, C.}
\newblock {A}daptive {S}pectral {E}lement {S}imulations of {T}hin {F}lame
  {S}heet {D}eformations.
\newblock {\em J. Sci. Comput. 17\/} (2002), 1--3.

\bibitem{Fischer:2002sx}
{\sc Fischer, P.~F., Kruse, G., and Loth, F.}
\newblock Spectral element methods for transitional flows in complex
  geometries.
\newblock {\em J. Sci. Comput. 17\/} (2002), 81--98.

\bibitem{Fournier:2004wk}
{\sc Fournier, A., Taylor, M.~A., and Tribbia, J.~J.}
\newblock The spectral element atmosphere model ({SEAM}): {H}igh-resolution
  parallel computation and localized resolution of regional dynamics.
\newblock {\em Mon. Wea. Rev. 132}, 3 (2004), 726--748.

\bibitem{Frohn:2002zp}
{\sc Frohn, L.~M., Christensen, J.~H., and Brandt, J.}
\newblock Development of a high-resolution nested air pollution model -- the
  numerical approach.
\newblock {\em J. Comput. Phys. 179\/} (2002), 68--94.

\bibitem{Galewsky:2004wa}
{\sc Galewsky, J., Scott, R.~K., and Polvani, L.~M.}
\newblock An initial-value problem for testing numerical models of the global
  shallow water equations.
\newblock {\em Tellus, A 56}, 5 (2004), 429--440.

\bibitem{Gibbs:1899fk}
{\sc Gibbs, J.~W.}
\newblock {F}ourier {S}eries.
\newblock {\em Nature 59\/} (1899), 606.

\bibitem{Giraldo:2000nu}
{\sc Giraldo, F.~X.}
\newblock The {L}agrange-{G}alerkin method for the two-dimensional shallow
  water equations on adaptive grids.
\newblock {\em Int. J. Numer. Meth. Fluids 33\/} (2000), 789--832.

\bibitem{Giraldo:2001ys}
{\sc Giraldo, F.~X.}
\newblock A spectral element shallow water model on spherical geodesic grids.
\newblock {\em Int. J. Numer. Meth. Fluids 35\/} (2001), 869--901.

\bibitem{Giraldo:2006gw}
{\sc Giraldo, F.~X.}
\newblock High-order triangle-based discontinuous {G}alerkin methods for
  hyperbolic equations on a rotating sphere.
\newblock {\em J. Comput. Phys 214\/} (2006), 447--465.

\bibitem{Giraldo:2005yj}
{\sc Giraldo, F.~X., and Warburton, T.}
\newblock A nodal triangle-based spectral element method for the shallow water
  equations on the sphere.
\newblock {\em J. Comput. Phys. 207}, 1 (2005), 129--150.

\bibitem{Gopalakrishnan:2002in}
{\sc Gopalakrishnan, S.~G., Bacon, D.~P., Ahmad, N.~N., Boybeyi, Z., Dunn,
  T.~J., Hall, M.~S., Jin, Y., Lee, P. C.~S., Mays, D.~E., Madala, R.~V.,
  Sarma, A., Turner, M.~D., and Wait, T.~R.}
\newblock An operational multiscale hurricane forecasting system.
\newblock {\em Mon. Wea. Rev. 130\/} (2002), 1830--1847.

\bibitem{Haidvogel:1997kx}
{\sc Haidvogel, D., Curchitser, E., Iskandarani, M., Hughes, R., and Taylor,
  M.}
\newblock Global modeling of the ocean and atmosphere using the spectral
  element method.
\newblock {\em Atmos.-Ocean 35\/} (1997), 505.

\bibitem{Iselin:2005ma}
{\sc Iselin, J., Gutowski, W.~J., and Prusa, J.~M.}
\newblock Tracer advection using dynamic grid adaptation and {MM5}.
\newblock {\em Mon. Wea. Rev. 133\/} (2005), 175--187.

\bibitem{Iselin:2002ff}
{\sc Iselin, J., Prusa, J.~M., and Gutowski, W.~J.}
\newblock Dynamic grid adaptation using the {MPDATA} scheme.
\newblock {\em Mon. Wea. Rev. 130\/} (2002), 1026--1039.

\bibitem{Jablonowski:2004fa}
{\sc Jablonowski, C.}
\newblock {\em Adaptive Grids in Weather and Climate Modeling}.
\newblock PhD thesis, University of Michigan, Ann Arbor, MI, 2004.
\newblock {D}epartment of Atmospheric, Oceanic and Space Sciences, 292 pp.

\bibitem{Jablonowski:2004}
{\sc Jablonowski, C., Herzog, M., Penner, J.~E., Oehmke, R.~C., Stout, Q.~F.,
  and van Leer, B.}
\newblock Adaptive grids for weather and climate models.
\newblock In {\em ECMWF Seminar Proceedings on {R}ecent {D}evelopments in
  {N}umerical {M}ethods for {A}tmosphere and {O}cean {M}odeling\/} ({R}eading,
  {U}nited {K}ingdom, September 2004), ECMWF, pp.~233--250.

\bibitem{Jablonowski:2006b}
{\sc Jablonowski, C., Herzog, M., Penner, J.~E., Oehmke, R.~C., Stout, Q.~F.,
  and van Leer, B.}
\newblock Adaptive grids for atmospheric general circulation models: Test of
  the dynamical core, 2006.
\newblock to be submitted to Mon. Wea. Rev.

\bibitem{Jablonowski:2006}
{\sc Jablonowski, C., Herzog, M., Penner, J.~E., Oehmke, R.~C., Stout, Q.~F.,
  van Leer, B., and Powell, K.~G.}
\newblock Block-structured adaptive grids on the sphere: Advection experiments.
\newblock {\em Mon. Wea. Rev. 134}, 12 (2006), 3691--3713.

\bibitem{jakob:93}
{\sc Jakob, R., Hack, J.~J., and Williamson, D.~L.}
\newblock Solutions to the shallow-water test set using the spectral transform
  method.
\newblock Tech. Rep. NCAR/TN-388+STR, National Center for Atmospheric Research,
  Boulder, Colorado, 1993.

\bibitem{jakob:95}
{\sc Jakob-Chien, R., Hack, J.~J., and Williamson, D.~L.}
\newblock Spectral transform solutions to the shallow water test set.
\newblock {\em Journal of Computational Physics 119\/} (1995), 164--187.

\bibitem{Karypis:2002yo}
{\sc Karypis, G., Schloegel, K., and Kumar, V.}
\newblock {\em {ParMetis: P}arallel Graph Partitioning and Sparse Matrix
  Ordering Library, Version 3.0}.
\newblock University of Minnesota, 2002.

\bibitem{Kurihara:1980jj}
{\sc Kurihara, Y., and Bender, M.~A.}
\newblock Use of a movable nested-mesh model for tracking a small vortex.
\newblock {\em Mon. Wea. Rev. 108\/} (1980), 1792--1809.

\bibitem{Kurihara:1979zv}
{\sc Kurihara, Y., Tripoli, G.~J., and Bender, M.~A.}
\newblock Design of a movable nested-mesh primitive equation model.
\newblock {\em Mon. Wea. Rev. 107\/} (1979), 239--249.

\bibitem{laeuter:2004}
{\sc L{\"a}uter, M.}
\newblock {\em Gro{\ss}r{\"a}umige {Z}irkulationsstrukturen in einem
  nichtlinearen adaptiven {A}tmosph{\"a}renmodell}.
\newblock PhD thesis, Universit{\"a}t Potsdam, Germany, 2004.
\newblock {W}issenschaftsdisziplin Physik der Atmoph{\"a}re, 135 pp.

\bibitem{laeuter:05}
{\sc L{\"a}uter, M., Handorf, D., and Dethloff, K.}
\newblock Unsteady analytical solutions of the spherical shallow water
  equations.
\newblock {\em J. Comput. Phys. 210\/} (2005), 535--553.

\bibitem{laeuter:2006}
{\sc L{\"a}uter, M., Handorf, D., Rakowsky, N., Behrens, J., Frickenhaus, S.,
  Best, M., Dethloff, K., and Hiller, W.}
\newblock A parallel adaptive barotropic model of the atmosphere, 2006.
\newblock J. Comput. Phys., accepted.

\bibitem{Lin:1996}
{\sc Lin, S.-J., and Rood, R.~B.}
\newblock Multidimensional flux-form semi-{L}agrangian transport scheme.
\newblock {\em Mon. Wea. Rev. 124\/} (September 1996), 2046--2070.

\bibitem{Lin:1997}
{\sc Lin, S.-J., and Rood, R.~B.}
\newblock An explicit flux-form semi-{L}agrangian shallow water model on the
  sphere.
\newblock {\em Quart. J. Roy. Meteor. Soc. 123\/} (1997), 2477--2498.

\bibitem{Mavriplis:1989xe}
{\sc Mavriplis, C.}
\newblock {\em Nonconforming Discretizations and a Posteriori Error Estimators
  for Adaptive Spectral Element Techniques}.
\newblock PhD thesis, MIT, Boston, MA., 1989.
\newblock 156 pp.

\bibitem{nair:02}
{\sc Nair, R.~D., and Machenhauer, B.}
\newblock The mass-conservative cell-integrated semi-{L}agrangian advection
  scheme on the sphere.
\newblock {\em Mon. Wea. Rev. 130\/} (March 2002), 649--667.

\bibitem{Nair:2005aj}
{\sc Nair, R.~D., Thomas, S.~J., and Loft, R.~D.}
\newblock A discontinuous {G}alerkin global shallow water model.
\newblock {\em Mon. Wea. Rev. 133}, 4 (2005), 876--888.

\bibitem{oehmke:04}
{\sc Oehmke, R.~C.}
\newblock {\em High Performance Dynamic Array Structures}.
\newblock PhD thesis, University of Michigan, Ann Arbor, MI, USA, 2004.
\newblock 93 pp.

\bibitem{oehmke:01}
{\sc Oehmke, R.~C., and Stout, Q.~F.}
\newblock Parallel adaptive blocks on a sphere.
\newblock In {\em Proc. 10th {SIAM} Conference on Parallel Processing for
  Scientific Computing\/} (Portsmouth, Virginia, USA, 2001), SIAM.
\newblock {CD-ROM}.

\bibitem{Patera:1984uq}
{\sc Patera, A.~T.}
\newblock A spectral element method for fluid dynamics: {L}aminar flow in a
  channel expansion.
\newblock {\em J. Comput. Phys. 54\/} (1984), 468--488.

\bibitem{Prusa:2003vs}
{\sc Prusa, J.~M., and Smolarkiewicz, P.~K.}
\newblock An all-scale anelastic model for geophysical flows: dynamic grid
  deformation.
\newblock {\em J. Comput. Phys. 190\/} (2003), 601--622.

\bibitem{Quarteroni:1997kw}
{\sc Quarteroni, A., and Valli, A.}
\newblock {\em Numerical Approximations of Partial Differential Equations},
  vol.~23 of {\em Springer series in computational mathematics}.
\newblock Springer-Verlag, 1997.

\bibitem{Rosenberg:2006zs}
{\sc Rosenberg, D., Fournier, A., Fischer, P., and Pouquet, A.}
\newblock Geophysical--astrophysical spectral-element adaptive refinement
  ({GASpAR}): {O}bject-oriented h-adaptive fluid dynamics simulation.
\newblock {\em J. Comput. Phys. 215\/} (2006), 59--80.

\bibitem{Sadourny:1972xo}
{\sc Sadourny, R.}
\newblock Conservative finite-difference approximations of the primitive
  equations on quasi-uniform spherical grids.
\newblock {\em Mon. Wea. Rev. 100\/} (1972), 136--144.

\bibitem{Skamarock:1989gy}
{\sc Skamarock, W.~C.}
\newblock Truncation error estimates for refinement criteria in nested and
  adaptive models.
\newblock {\em Mon. Wea. Rev. 117\/} (1989), 872--886.

\bibitem{Skamarock:1993mk}
{\sc Skamarock, W.~C., and Klemp, J.~B.}
\newblock Adaptive grid refinement for twodimensional and three-dimensional
  nonhydrostatic atmospheric flow.
\newblock {\em Mon. Wea. Rev. 121\/} (1993), 788--804.

\bibitem{Skamarock:1989ya}
{\sc Skamarock, W.~C., Oliger, J., and Street, R.~L.}
\newblock Adaptive grid refinement for numerical weather prediction.
\newblock {\em J. Comput. Phys. 80\/} (1989), 27--60.

\bibitem{Smolarkiewicz:2006ko}
{\sc Smolarkiewicz, P.~K.}
\newblock Multidimensional positive definite advection transport algorithm: an
  overview.
\newblock {\em Int. J. Num. Meth. Fluids 50}, 10 (2006), 1123--1144.

\bibitem{Smolarkiewicz:2005ih}
{\sc Smolarkiewicz, P.~K., and Szmelter, J.}
\newblock {MPDATA:} an edge-based unstructured-grid formulation.
\newblock {\em J. Comput. Phys. 206}, 2 (2005), 624--649.

\bibitem{St-Cyr:2006xf}
{\sc St-Cyr, A., Dennis, J.~M., Thomas, S.~J., and Tufo, H.~M.}
\newblock Operator splitting in a non-conforming adaptive spectral element
  atmospheric model.
\newblock {SIAM J. S}ci. Comput., submitted, 2006.

\bibitem{St-Cyr:2005ur}
{\sc St-Cyr, A., and Thomas, S.~J.}
\newblock Nonlinear {OIFS} for a hybrid {G}alerkin atmospheric model.
\newblock In {\em Lecture notes in computer science\/} (2005), vol.~LNCS 3516
  of {\em Lecture Notes in Computer Science}, pp.~57--63.

\bibitem{St-Cyr:2005jw}
{\sc St-Cyr, A., and Thomas, S.~J.}
\newblock Nonlinear operator-integration factor splitting for the shallow water
  equations.
\newblock {\em Appl. Numer. Math. 52\/} (2005), 429--448.

\bibitem{Taylor:1997vn}
{\sc Taylor, M., Tribbia, J., and Iskandarani, M.}
\newblock The spectral element method for the shallow water equations on the
  sphere.
\newblock {\em J. Comput. Phys. 130}, 1 (1997), 92--108.

\bibitem{Thomas:2002ti}
{\sc Thomas, S.~J., and Loft, R.~D.}
\newblock Semi-implicit spectral element atmospheric model.
\newblock {\em J. Sci. Comp. 17\/} (2002), 339--350.

\bibitem{tolstykh:02}
{\sc Tolstykh, M.~A.}
\newblock Vorticity-divergence semi-{L}agrangian shallow-water model of the
  sphere based on compact finite differences.
\newblock {\em J. Comput. Phys. 179\/} (2002), 180--200.

\bibitem{Tufo:1998ey}
{\sc Tufo, H.~M.}
\newblock {\em Algorithms for large-scale parallel simulations of unsteady
  incompressible flows in three-dimensional complex geometries}.
\newblock PhD thesis, Division of Applied Mathematics, Brown University,
  Providence, R.I., 1998.

\bibitem{Vandeven:1991ef}
{\sc Vandeven, H.}
\newblock Family of spectral filters for discontinuous problems.
\newblock {\em J. Sci. Comput. 6\/} (1991), 159--192.

\bibitem{Wilhelm:uo}
{\sc Wilhelm, D., and Kleiser, L.}
\newblock Stable and unstable formulation of the convection operator in
  spectral element simulations.
\newblock {\em Appl. Numer. Math. 33\/} (2000), 275--280.

\bibitem{Williamson:1992ts}
{\sc Williamson, D.~L., Drake, J.~B., Hack, J.~J., Jakob, R., and Swarztrauber,
  P.~N.}
\newblock A standard test set for numerical approximations to the shallow water
  equations in spherical geometry.
\newblock {\em J. Comp. Phys. 102\/} (1992), 211--224.

\end{thebibliography}

\end{document}